\documentclass[aps]{revtex4}

\usepackage{graphicx}
\usepackage{amsmath}
\usepackage{amsfonts}
\usepackage{mathrsfs}
\usepackage{color}
\usepackage{slashed}
\usepackage{stackrel}
\newcommand{\tn}{\textbf}

\def\<{\langle}
\def\>{\rangle}

\def\beq{\begin{equation}}
\def\eeq{\end{equation}}
\def\barray{\begin{eqnarray}}
\def\earray{\end{eqnarray}}

\newcommand{\be}{\begin{equation}}
\newcommand{\ee}{\end{equation}}
\def\ba{\begin{eqnarray}}
\def\ea{\end{eqnarray}}

\font\numbers=cmss12
\font\upright=cmu10 scaled\magstep1
\def\stroke{\vrule height8pt width0.4pt depth-0.1pt}
\def\topfleck{\vrule height8pt width0.5pt depth-5.9pt}
\def\botfleck{\vrule height2pt width0.5pt depth0.1pt}
\def\Zmath{\vcenter{\hbox{\numbers\rlap{\rlap{Z}\kern
0.8pt\topfleck}\kern 2.2pt
                   \rlap Z\kern 6pt\botfleck\kern 1pt}}}
\def\Qmath{\vcenter{\hbox{\upright\rlap{\rlap{Q}\kern
                   3.8pt\stroke}\phantom{Q}}}}
\def\Nmath{\vcenter{\hbox{\upright\rlap{I}\kern 1.7pt N}}}
\def\Rmath{\vcenter{\hbox{\upright\rlap{I}\kern 1.5pt R}}}
\def\Pmath{\vcenter{\hbox{\upright\rlap{I}\kern 1.5pt P}}}
\def\Cmath{\slashed \Pmath}

\begin{document}

\title{There is entanglement in the  primes }
\author{Jos\'e I. Latorre$^1$ and Germ\'an  Sierra$^2$ \vspace{0.2cm} \\  
${}^1$ Departament d'Estructura i Constituents de la Mat\`eria, Universitat de Barcelona, 
Barcelona, Spain,   \\ 
Centre for Quantum Technologies, National University of Singapore, Singapore.\\
 ${}^2$  Instituto de F\'isica Te\'orica UAM/CSIC, Universidad Aut\'onoma de Madrid, Cantoblanco, 
 Madrid, Spain.}

\begin{abstract}
Large series of prime numbers can be superposed on a single
quantum register and then analyzed in full parallelism. 
The construction of this Prime state is efficient,
as it hinges on the use of a quantum version of any efficient
primality test. We show that the Prime state turns out to be very
entangled as shown by the scaling properties of purity, Renyi entropy and
von Neumann entropy. An analytical approximation to these measures of entanglement can be obtained from the detailed analysis of the  
entanglement spectrum of the Prime state, which in turn produces
new insights in the Hardy-Littlewood conjecture for the pairwise distribution of  primes.
The extension of these ideas to a Twin Prime state shows that this new
state is even more entangled than the Prime state, obeying
majorization relations. We further discuss the construction of quantum
states that encompass relevant series of numbers and opens the
possibility of applying quantum computation to Arithmetics in novel
ways.
\end{abstract}

\maketitle

\section{Introduction} 

Prime numbers have fascinated mathematicians and physicist for  centuries.
They are the  building blocks in Number Theory that explains its
relevance for Pure Mathematics 
\cite{Apostol,D}. 
However, we do  not  know of any fundamental physical
theory that is  based on deep facts in Number Theory \cite{V}.  
In spite of this,  there have been several attempts in the past to provide a physical meaning to  prime numbers, 
with the hope that both fields can benefit from each other. Without trying to be exhaustive, we shall mention
some of these attempts that will also aid us to frame  the aims   of this work (for a review on this topic
see \cite{SH11} and the database \cite{Wat}). 

 Prime numbers in Physics
have been considered as classical or quantum objects. In the first category falls  the  gas of primons 
where the particles have energies proportional to the logarithm of the primes,
and whose partition function is the Riemann zeta function $\zeta(s)$, with $s$ a real number  related to 
the temperature of the gas \cite{J90}-\cite{BB91}. The divergence of $\zeta(s)$ at $s=1$ being related to a Hagedorn transition
that also appears  in string theory \cite{H68}.  Another example of a classical
interpretation of the primes appears in the spectral approach to the Riemann Hypothesis. 
According to Berry and collaborators,  the 
primes correspond to the primitive periodic orbits of a classical  chaotic Hamiltonian whose quantization
would yield  the Riemann zeros as eigenenergies \cite{B86}-\cite{B03}. This would provide a proof
of the celebrated Riemann Hypothesis, that as Berry put it, 
would mean that ``there is music in the primes"  \cite{B12}-\cite{B92}. 
This conjecture  has been very fruitful in the past, helping 
to  establish many links   between  Number Theory, Quantum Chaos and Random Matrix Theory,  the later theory
being invented in Nuclear Physics in order to understand the spectrum of complex nuclei, and
that later on found numerous  applications in Condensed Matter Physics \cite{M73}-\cite{Me}. This line
of thought runs parallel with several physical approaches to give a physical realization of the Riemann zeros \cite{BK99b}-\cite{N14}.

A quantum interpretation of the primes was given by Mussardo, who constructed a Hamiltonian
whose spectrum  coincides in average with the position of the 
prime numbers \cite{M97}. Later numerical attempts to match the precise positions of the primes, gave a potential
with a multi fractal structure. 
A similar idea concerning the Riemann zeros,  was  pursued by Wu and Sprung \cite{WS},  and continued by several groups,
finding a fractal  potential whose  fractal dimension is smaller than the one associated to the prime potential  \cite{R95}-\cite{SBH08} 
(see \cite{SH11} for a review). 
These results seem quite natural  given that the primes satisfy an almost Poisson distribution, while the Riemann
zeros  follow the statistics of the Gaussian Unitary Ensemble (GUE) of Random Matrix Theory \cite{M73}-\cite{OO}. 

Another quantum realization of the primes  was  proposed by the authors of this paper using a quantum computer \cite{LS}.
Rather than dealing with  the primes individually, they are taken as a whole using  the available computational resources. 
The benefit of this realization is that number theoretical functions, such as the prime counting function \cite{E74}, 
become accessible through measurement. In this manner one can verify,  but not prove, the Riemann Hypothesis
beyond the current classical methods. This method provides a quantum speed up of classical algorithms,
which is analogous to the well known Shor's algorithm of factorization \cite{S97},  although the speed in our case
is not exponential but polynomial since it is based in the Grover's quantum search algorithm \cite{G96}. In both methods, the  origin of  
the quantum gain is entanglement,  that  is the central concept in Quantum Information Theory \cite{NC00}.  In this paper 
we shall apply  standard techniques to analyze the entanglement hidden in the Prime state, such as
von Neumann entropy, Renyi entropies, etc. We mentioned above that the primes are almost random numbers. 
One would then have  expected that the Prime  state would  also be random with the highest
von Neumann entropy for bipartition's. However, this is not the case. The prime state is certainly random
but not a typical random state in the Hilbert space. The  von Neumann entropy grows linearly with the number
of qubits, so it violates  the so called area law that  characterizes the  low energy states of local Hamiltonians \cite{A08}-\cite{M09}. 
But the coefficient of the linear growth is smaller than the one of a generic random state, a property
that is relevant to  black-hole evaporation
\cite{P93,P93b}. This result
indicates the existence of correlations built in  the Prime state. That the  prime numbers are correlated
was  famously conjectured long ago  by Hardy and Littlewood, who  found an heuristic
formula for the probability of  having pairs of primes  $p$ and $p+k$ \cite{HL}.  Quite interestingly, 
we have found  that the Hardy-Littlewood pair distribution law is what characterizes 
the bipartite entanglement properties  of the Prime state.  Hence the entanglement measures,
such as the von Neumann entropy or Renyi entropies, can be computed from the pair
correlations of primes for a large number of qubits. It is worthwhile to mention that
the Hardy-Littlewood law played an important role in the derivation of the GUE statistics
of the Riemann zeros that was first conjectured by Montgomery and confirmed numerical by Odlyzko (for a review see \cite{K93,BK99}).  
In our case, the computation of the Renyi entropies involves averages of the Hardy-Littlewood  law, 
that suggests  possible links between the two approaches, namely  treating the primes
as quantum objects or as classical ones. Other physical oriented approaches to prime numbers 
can be found in references  \cite{B73}-\cite{GS14}. 

The plan  of the paper is as follows. In Sect. II we review briefly the Prime state 
proposed in reference \cite{LS}. In Sect. III we compute the entanglement figures of merit for the Prime state. An analytical approximation to the entanglement spectrum is proposed and exploited in Sect. IV. In Sect. V, we define new number theoretical states, that generalize
 the construction of the prime state. We have included four appendices with the more technical
 details.

\section{Review of the prime state}

The key idea to address prime numbers using a quantum computer consists in 
exploiting the quantum superposition principle. To be more precise, we shall
analyze  the possibility of 
producing a single quantum state made of the superposition of prime numbers. In this way, all prime numbers can be manipulated in full parallelism as a single entity bringing the possibility of exploring their properties in a much more efficient way than in classical computations. In this section we shall review this idea as  presented in \cite{LS}.

The central object of our analysis is the Prime state $|\Pmath_n\rangle$, which corresponds to a quantum state made of the equally weighted superposition of
prime numbers as written in the $n$-qubit computational basis \cite{LS}
\beq 
| \Pmath_n \rangle = \frac{ 1}{ \sqrt{\pi(2^n)} } \sum_{p\in {\rm prime}  < 2^n}  |p \rangle,
\label{1}
\eeq
where each prime number $p=p_0 2^0+p_1 2^1+\ldots + p_{n-1} 2^{n-1}$ is
implemented as a ket
$|p \rangle= |p_{n-1}, \dots, p_0 \rangle$,
$N = 2^n$ and $\pi(N)$ is 
a constant that normalizes the state. As an example, we can write 
explicitly the Prime state for 4 qubits
\begin{eqnarray}
| \Pmath_4 \rangle &= &\frac{ 1}{ \sqrt{\pi(16)} }
\left( |2\rangle+ |3\rangle+ |5\rangle+ |7\rangle+ |11\rangle+ |13\rangle
\right) \\
&=&\frac{ 1}{\sqrt 6 }
\left( |0010\rangle+ |0011\rangle+ |0101\rangle+ |0111\rangle+ |1011\rangle+ |1101\rangle
\right) .
\label{example}
\end{eqnarray}
Note that 100 qubits would be sufficient to superpose far more primes than
the ones  that can be analyzed using a classical computer at present.

The normalization of the Prime state exactly corresponds to the total number
of primes below $N$, that is the fundamental prime number counting function $\pi(N)$. The Prime Number Theorem (PNT) provides an estimate for this quantity in terms of the Log integral function \cite{E74}
\beq
\pi(N)  \sim Li(N) \overset{N \to \infty}{\longrightarrow} 
 \frac{N}{\ln N} .
\label{PNT}
\eeq 
The actual count of prime numbers does necessarily fluctuate around the PNT estimation,
as Littlewood \cite{Li14} proved that $\pi(N)-Li(N)$ will change sign infinitely many times. 
Moreover, the fluctuations of the actual $\pi(N)$ around
the estimate $Li(N)$ are bounded for large $N$ if the Riemann Hypothesis holds true \cite{K01,S76}
\beq
\left| \pi(N)-Li(N) \right| \le \frac{1}{8\pi}{\sqrt N} \ln N \ .
\label{Riemannbound}
\eeq
It is then possible to falsify the  Riemann Hypothesis if the actual number
of prime numbers would be proven to depart from the PNT estimation beyond the bound in Eq. (\ref{Riemannbound}).

\subsection{ Construction of the Prime state}

A first and critical question is whether the Prime state can in principle be constructed and, if so, whether its construction is efficient in the sense that it uses resources in space and time which only grow  polynomially with the number of bits of the primes under consideration. There are two strategies to construct the Prime state that we shall briefly present.

 Let us first
focus on a probabilistic procedure to construct the Prime state that hinges on the action of a primality check operator.  At the outset the register must be initialized as a product state of all qubits in their 0 logical value. Then a 
 set of Hadamard gates acting on each qubit $U_H | 0\rangle=\frac{1}{2}(|0\rangle+ |1\rangle )$ generates
the global superposition
\begin{equation}
U^{(0)}_H \otimes \dots \otimes  U^{(n-1)}_H |00\ldots 0\rangle=
\frac{1}{\sqrt {2^n}} \sum_{x=0}^{2^n-1} |x\rangle, 
\end{equation}
where $|x\rangle =|x_{n-1}, \ldots,x_{1}, x_0\rangle$.
 The key step in the construction of
the Prime state is to use a
 unitary operator that discriminates primes from composites. The detailed
 discussion of this primality check operator will come later. Here it suffices to
consider that there exists an efficient quantum circuit such that
\begin{equation}
  U_{\rm {primality}} 
  |x\rangle |0\rangle=\left\{
  \begin{array}{c l}
    |x\rangle |0\rangle & x\in {\rm primes}\\
    |x\rangle |1\rangle & x\notin {\rm primes}\\
  \end{array} \right. .
\end{equation}
The action of this unitary operation on the superposition of all states reads
\begin{equation}
  U_{\rm {primality}} \frac{1}{\sqrt 2^n}\sum_{x=0}^{2^n-1}|x\rangle |0\rangle=
  \frac{1}{\sqrt {2^n}}
  \left(  \sqrt{\pi(2^n)}  |\Pmath_n\rangle |0\rangle+
\sqrt{ 2^n - \pi(2^n)} |\Cmath_n\rangle |1\rangle\right) \ ,
\end{equation}
where $|\Cmath_n\rangle$ is a normalized quantum state made out of the superposition
of all composite numbers less or equal to $2^n-1$.
The Prime state is obtained when a measurement of the ancilla produces
the 0 readout. The results of quantum measurements are in general probabilistic. In our case, the Prime state is obtained upon measurement with probability
\begin{equation} 
  {\rm Prob}(\Pmath_n)= \frac{\pi(2^n)}{2^n}\sim \frac{1}{n \ln 2} \ ,
\end{equation}
where we have use the PNT.
This construction shows that obtaining the Prime state is an efficient task.
To be precise, using the standard amplification technique the number of repetitions for this construction to produce an instance of the Prime state with probability exponentially close to 1
only grows polynomially with the number of qubits $n$.

There is a second, deterministic procedure to construct the Prime state.  
The basic ingredient remains the use of a primality operator with the difference that such operator is called  an oracle
and acts as 
\begin{equation}
U_{\rm oracle} |x\rangle = (-1)^{\chi_\pi(x)} |x\rangle \ ,
\label{oracle}
\end{equation}
where $\chi_\pi(x)=1$ for prime numbers and 0 otherwise ($\chi_\pi(x)$ is the characteristic function for primes). 
That is,  $U_{\rm oracle} $ corresponds to a primality quantum oracle that can act on a superposition of states changing the sign of only prime states. Following the standard analysis of Grover's algorithm, the Prime state can be obtained  using
$k$ calls to the oracle, where in our case
\beq
k=\frac{\pi}{4}\sqrt{\frac{N}{M}}\sim \sqrt{n} \ .
\eeq
where $M$ is the number of states that change sign upon the action of the
oracle. In our case $M=\pi(N)$, thus $\sqrt {N/M}\sim {\sqrt n}$. 
Grover's construction is particularly powerful, as a Prime state of 100 qubits would only need 7 calls to the
quantum primality oracle. 

Both construction techniques, the first based on probabilistic measurements and the second on deterministic action of an oracle, show that the Prime state can be efficiently created on a quantum computer.

\subsection{Primality quantum oracle}

Both the probabilistic and the deterministic constructions of the Prime state make use of the possibility of checking primality on a superposition   states. This is tantamount to transform a classical primality test into a
quantum circuit. It is very fortunate that primality check was proven
to be in class P \cite{AKS}. This means that deciding whether a number is prime 
or not can be determined efficiently on a classical computer, as proven in Ref. 
\cite{AKS}. 

An explicit circuit for a primality quantum oracle was proposed in Ref. \cite{LS}. There, rather than
using the intricate algorithm from \cite{AKS}, it was proposed to 
create a quantum version of the Rabin-Miller primality test \cite{MR}. Such a test works as follows. Let us consider the problem of determining whether
$x$ is prime or composite. First we write $x-1=d \ 2^s$, where $d$ is odd.
We then choose a witness $a$, $1 \le a \le x$ and compute
\begin{eqnarray}
 &&a^d\not \equiv  1\ \pmod x \quad   \label{MR} \\
 &&a^{2^r d} \not \equiv -1 \pmod x\qquad 0\le r\le s-1. \nonumber 
\end{eqnarray}
If all these tests are verified, $x$ is declared composite. 
However if any of these tests fails, then $x$ can be either prime or composite so that no decision can be taken. The witness
$a$ is then called a strong liar to $x$. The difficulty to use this kind of test
emerges from the fact that prime numbers are not certified with certainty but only
probabilistically.
The way to solve the problem of strong liars is to repeat the test with  different witnesses, so that  the probability to be deceived by strong liars vanishes. 
Assuming the Generalized Riemann Hypothesis, the Miller-Rabin test is deterministic using less than $\log^2 x$ witnesses. 
For instance, all numbers below $x<3\ 10^{14}$ can be correctly classified as prime or composite using as witnesses $a=2,3,5,7,11,13,17$.

The detailed construction of a quantum oracle based on the primality Rabin-Miller 
test is presented in Ref. \cite{LS}. The computational cost of the oracle only grows as $n^6$, which remains polinomial and, thus, efficient.   

\subsection{Verification of Riemann Hypothesis}

The algorithm we have presented to construct the Prime state using a primality quantum oracle can be modified to perform a verification of Riemann Hypothesis. The idea is to entangle the calls to the oracle with a set of ancillas. Then, a Quantum Fourier transform is performed on these ancillas
providing a measurement of the total amount of states that verify the 
oracle condition. The structure of this Quantum Counting algorithm was
introduced in Ref. \cite{BHT}.

We here need not repeat the details of the Quantum Counting algorithm, it is  enough to recall the main result. We are interested in counting the number of
solutions of the primality oracle, that is
 $M=\pi(N)$, with a bounded error. The Quantum Counting algorithm produces an estimate $\tilde M=\pi_{QM}(N)$ to the actual number of solutions $M$ to the oracle such that
\be
  \left|\tilde M- M\right|< \frac{2 \pi}{c}  M^{1/2}+ \frac{\pi^2}{c^2}, 
\ee
where  $c$ is a constant, 
using only $c \,  N^{1/2}$  calls to the oracle. 
The relevant fact is that using the PNT our quantum algorithm can verify the
prime counting function with an accuracy
\be
   \left|\pi_{QM}(N)-\pi(N)\right|< \frac{2 \pi}{c}  \frac{N^{1/2}}{\log^{1/2} N}
\label{quantumestimate2}
\ee
where $\pi_{QM}(N)$ is the result of the quantum computation which only needs $c\sqrt{N}$ calls
to the oracle
and uses  $O(n=\log(N))$ space allocation.  Note that the accuracy needed for the verification of the
Riemann Hypothesis should be smaller than  $N^{1/2} \log   N$. Therefore, quantum counting on the Prime state is indeed sufficient
to verify departures from the Riemann Hypothesis.

Let us now show that our quantum counting algorithm outperforms 
other classical strategies. First of all, let us consider a brute force 
random generator of numbers checked by classical primality checks. Successive use of this strategy
would deliver an estimator $\hat\pi(N)$ for the real $\pi(N)$ and an estimator for its variance $\hat {\rm Var}$.
To first order the observed distribution of primes among all numbers can be thought as a binomial
distribution, giving the output 0  for primes and 1 for composites, with a binomial probability for
the former to be $p=\pi(N)/N\sim 1/n$. This approximate binomial distribution
should be corrected for the fact that primes get scarcer as the numbers grow. 
Nevertheless we can go ahead with this approach, an try an estimator $\hat p$ of  $p$ checking
primality on $k$ numbers smaller than $N$. Then the estimator  $\hat p$ is made out of 
the total number of primes $k_p$ found in the set of $k$ numbers,
$\hat p= k_p/k$. The relevant point is that the estimator for the variance is
\be
\hat {Var}= \frac{\hat p(1-\hat p)}{k}=\frac{k_p(1-k_p/k)}{k^2} .
\ee
Thus, this classical randomized strategy would beat the precision needed for verifying
the Riemann Hypothesis if $N$ numbers are tried, so that
\be
\hat p\pm \sqrt{\frac{\hat p(1-\hat p)}{k}} \overset{k=N}{\longrightarrow} 
\hat \pi(N)\pm \sqrt{\frac{N}{\log N}}
\ee
Yet, in the case where only $k\sim \sqrt{N}$ calls are made, the reduction of the variance is softer, leaving
an error of the order of $N^{3/4}/ \log N$, which is insufficient to verify the Riemann Hypothesis. 
This shows that quantum counting speeds up the
computation of the fluctuations of $\pi(N)$ around $Li(N)$.

We can also compare the quantum counting approach to the best known 
 classical algorithms to obtain $\pi(N)$  as proposed by 
Lagarias, Miller, and  Odlyzko \cite{LMO}. This algorithm uses
 $N^{2/3}$ bit operations and the storage needed grows as
 $N^{1/3}$, where both scaling costs have log corrections. Lagarias and Odlyzko
have also proposed two  analytic $\pi(N)$-algorithms based on the Riemann zeta
function, whose order in  time and space  are $N^{ 3/5 + \epsilon} \; (\epsilon >0)$ and  $N^{\epsilon}$
in one case,  and $N^{ 1/2 + \epsilon}$ and  $N^{1/4+\epsilon}$ in the other case \cite{LO}, but none of these algorithms have been implemented numerically.  Classically, it is possible to 
find other algorithms that trade space with time, yet the product of time and memory is always
bigger than order $N^{1/2}$.  Hence the power of the quantum counting $\pi_{QM}(N)$,  given by Eq.(\ref{quantumestimate2}).

\subsection{Determination of Skewes number}

As far as exact computations have been performed, $Li(N)> \pi(N)$ for $N<10^{25}$ (see \cite{P12}). 
It is known that $Li(N)-\pi(N)$ must change sign infinitely many times,
but it is not known at which value of $N_0$ the first change of sign
will take place. 
A bound on this quantity is $N_0\sim e^{729}$,
which is called Skewes number. Classical computers can handle less than a hundred bits,
so they cannot address this problem.
 
Let us now argue that a quantum computer  could bound the position
of the change of sign if the fluctuation which is taking place is large enough.  We can again resort to the Quantum Counting algorithm
and explore the possibility to play with the relation between its precision and the goal of the computation. In the case of determining Skewes number, it is
necessary to measure the sign of $Li(N)-\pi(N)$. As an example, let us consider that a fluctuation has made $\pi(N)-Li(N)> {\sqrt N}$, that is the sign has
changed and a large fluctuation is present. Then, the algorithm presented above
is sufficient to determine this fluctuation since the expected accuracy goes as $\sqrt{ N/ \ln N}$ using $\sqrt(N)$ calls to the oracle. If fluctuations are smaller,
the original quantum counting algorithm can be adjusted. Indeed to obtain
an increasing accuracy 
\be 
| \pi_{QM}(N)-\pi(N)| <\frac{2 \pi}{c}\sqrt{\pi_{QM}(N)} + \frac{\pi^2}{c^2}, 
\ee
it is necessary to make a total of $c{\sqrt N}$ calls to the oracle.
The larger is $c$, the better the distance of $Li(N)$ to $\pi(N)$
can be discriminated. 

\subsection{Twin primes}

The Prime state can be used to generate other derived states of
relevance. As an instance, let us consider the case of twin primes.
A state made with the superposition of twin primes is easily
created using a measurement procedure as the one sketch
for the construction of the Prime state itself.
We first need to create a Prime state and, then, a global addition
of 2 is made on the register
\be
U_{+2} |\Pmath(n)\rangle = \frac{1}{\sqrt {\pi(2^n)}} \sum_{p \in {\rm primes}} | p+2\rangle. 
\ee
We then act with the primality operator
\begin{eqnarray}
 &&  U_{primality} \frac{1}{\sqrt{ \pi(2^ n)}}
 \sum_{p \in {\rm primes}} |p+2\rangle |0 \rangle \\ 
 \nonumber 
= &&
\frac{1}{\sqrt {\pi(2^ n)}}
\left( \sum_{(p,p+2)\in {\rm primes}}|p+2\rangle  |0\rangle
+\sum_{ p+2 \in {\rm composite}}
|p+2\rangle  |1\rangle\right).
\end{eqnarray}
A measurement on the ancilla with result 0 will essentially deliver the Twin Prime state (only substracting 2 to every state is necessary)
\be
|\Pmath_2(n)\rangle=
\frac{1}{\sqrt {\pi_2(2^ n)}}\sum_{(p,p+2)\in primes}|p\rangle , 
\ee
where $\pi_2(2^ n)$ is the counting function of twin primes below $2^n$.
The probability for the whole algorithm to produce the Twin Prime state 
from the original product state is $1/n^2$,  according to the conjecture
by Hardy-Littlewood \cite{HL},  which remains an efficient process. 

Similar algorithms  can be put forward to create superpositions of 
any sub-series of primes, including those of the form $p=a+b k$,
$(p,p+2,p+6)$, etc, as described in subsection III.B. All these 
constructions can be also viewed as slight modifications of the original
algorithm for the Prime state and remain efficient.

\section{Entanglement of the Prime state}

The entanglement properties of the Prime state can be analyzed considering
partitions of the quantum register in two sets $A$ and $B$. Then, 
the correlations between the two subsystems $A$ and $B$ can be 
quantified using different figures of merit. Here, we shall consider
measures of correlations such as purity and entropies of entanglement that we
shall first define. Genuine multipartite entanglement measures are
not available for states with a large number of qubits.

\subsection{Figures of merit of entanglement}

To study the entanglement of the prime state in Eq. (\ref{1}) we split 
the $n \geq 2$ qubits into two  disjoint sets $A$ and $B$,  where $A$ contains
the lowest $m$ qubits and $B$ the remaining $m'=n - m$ qubits 
(we take $m \in [1, n-1]$, so that $A$ and $B$ are not empty sets). 
This partition amounts to  the   decomposition of a prime number $p$ as

\ba
p & = &   a + 2^m b,  \label{2} \\
a & = &  p_0 2^0 + p_1 2^1 + \dots p_{m-1} 2^{m-1}, \qquad 
b =   p_{m}  2^0 + p_{m+1}  2^1 + \dots p_{n-1} 2^{m'-1}, \nonumber 
\ea
and correspondingly, the state in Eq. (\ref{1})  becomes 

\ba
| \Pmath_n \rangle  & = &  \frac{ 1}{ \sqrt{\pi(N)} } \sum_{a < 2^m, b    < 2^{m'}} \psi_{b,a}   \;  |b \rangle \otimes |a \rangle,
\label{3} \\
\psi_{b,a} & = &   \left\{ 
\begin{array}{cl} 
1,  & {\rm if} \, a + 2^m b : {\rm prime} \\
0,  & {\rm else} \\
\end{array}. 
\right.   \nonumber 
\ea
The  density matrix associated to the pure state in Eq. (\ref{3}) is  the projector

\ba
| \Pmath_n \rangle \langle \Pmath_n |  & = &  \frac{ 1}{\pi(N)}  \sum_{a, a'  < 2^m, b, b'     < 2^{m'}} \psi_{b,a} \psi_{b',a'} 
   \;  |b \rangle  \langle b' |  \otimes |a \rangle \langle a' |. 
\label{4} 
\ea
The reduced density matrices $\rho_A$ and $\rho_B$ are defined as 
the traces of Eq. (\ref{4}) over the  complementary Hilbert subspaces, namely

\ba
\rho_A = {\rm Tr}_B \;  | \Pmath_n \rangle \langle \Pmath_n |  & = & \sum_{a, a'  < 2^m} \rho^A_{a,a'}  
   \;  |a \rangle \langle a' |,
\label{5} \\   
\rho_B = {\rm Tr}_A \;  | \Pmath_n \rangle \langle \Pmath_n |  & = & \sum_{b, b'  < 2^{m'}} \rho^B_{b,b'}  
   \;  |b \rangle \langle b' |. \nonumber 
\ea
$\rho_A$ and $\rho_B$ and real symmetric matrices and satisfy the normalization conditions 
${\rm Tr}_{A} \; \rho_{A}  ={\rm Tr}_{B} \; \rho_{B}= 1$, that come from the normalization of the original
state in Eq. (\ref{3}).  
The corresponding  matrix elements are given by

\beq
\rho^A_{a, a'}     =   \frac{1}{ \pi(N) }  \sum_{b< 2^{m'}} \psi_{b,a} \, \psi_{b, a'},  \qquad 
\rho^B_{b, b'}     =    \frac{1}{ \pi(N) } \sum_{a< 2^{m}} \psi_{b,a} \, \psi_{b', a}. 
\label{6}  
\eeq
A generic property of reduced density matrices 
$\rho^ A$ and $\rho^ B$ constructed  from a pure state  
is that their  eigenvalues  $\lambda_i$ are the same (in the case where $\rho_A$ and $\rho_B$
have different dimensions, the latter statement holds for the non zero eigenvalues). 
The set  $\lambda_i$,  fully characterizes the bi-partite entanglement
between the blocks $A$ and $B$,  and satisfy the conditions $\lambda_i \in [0, 1]$ and 
$\sum_i \lambda_i =1$, as follows from the properties of $\rho_{A,B}$.
A figure of merit of entanglement is given by the von Neumann
entropy (or entanglement entropy) of the density matrix  $\rho_A$ (or $\rho_B$) 

\beq
S_A = - {\rm Tr}_A \; \rho_A \, \log \, \rho_A = - \sum_i \lambda_i \, \log \, \lambda_i. 
\label{7}
\eeq
Other measures of entanglement are provided by the Renyi entropies 

\beq
S_A^{(n)}  =  \frac{1}{1-n}  \log  {\rm Tr}_A \rho_A^n  = \frac{1}{1-n}  \log \sum_i \lambda_i^n, 
\label{8}
\eeq
where $n$ is a positive integer. In some cases, as in Conformal Field Theory,
one can compute  $S_A^{(n)}$ by a replica trick and make an analytic extension
in $n$, so that the entanglement entropy can be obtained in the limit  \cite{W94,V03,CC}

\beq
S_A = \lim_{n \rightarrow 1} S^{(n)}_A.
\label{9}
\eeq 
The entanglement entropy is a most relevant quantity that characterizes
bipartite entanglement, but a more detailed characterization is encoded
in the set of $\lambda_i's$, or alternatively in the so called entanglement 
spectrum $\varepsilon_i = - \log \lambda_i$, that can be thought of as 
fictitious eigenenergies of a Hamiltonian $H_E$ defined as $\rho_A = e^{ - H_E}$ \cite{LH}.

Eqs. (\ref{3}) and (\ref{6}) show that the density matrices $\rho_{A,B}$ 
should contain information about the way prime numbers are built up
and about their correlations. To   give their explicit expression 
we need to  introduce several  number theoretical functions.

\subsection{Review of prime number functions and theorems}

Let us now review some relevant functions that appear along the study
of prime numbers, as the entanglement properties of the Prime state
will end up relating them.
Consider an arithmetic progression   of the form $ a n + b \; (n=0, 1, \dots)$, where $a$ and $b$ 
are coprime numbers, that is, their greatest common divisor is 1 $(gcd(a,b)=1)$.
A famous theorem due to Dirichlet states the existence of  an infinite number of primes
for  each of these progressions \cite{Apostol,D}. The number of primes of the form $a n + b$, with $a$ and $b$
coprimes,  is denoted as $\pi_{a,b}(x)$, 

\beq
\pi_{a,b}(x) = \# \{ p=  a n + b \leq x, \;  p: {\rm prime},    \; gcd(a,b) = 1  \}.
\label{19b}
\eeq
The number of arithmetic progressions of this form is given by the Euler totient
function  $\phi(a)$ that  is  the number of coprime divisors of $a$, that is, the $b$'s  in Eq. (\ref{19b}). 
The counting function $\pi_{a,b}(x)$, satisfies a version of the PNT  given by 

\beq
\pi_{a,b}(x)  \simeq \frac{1}{\phi(a) } Li(x) \rightarrow  \frac{1}{\phi(a) }  \frac{x}{\log x}, \qquad \;  x \rightarrow \infty, 
\label{32}
\eeq
which  means  that the prime numbers are equally distributed, in average, among the progressions
$a n +b$. As an example, take $a=4$, whose coprime divisors are 1 and 3, that is $\phi(4) =2$. 
Hence, Eq.(\ref{32}) implies that the series $4 n +1$ and $4 n +3$ contain, asymptotically,  
half of the total number of primes. 

The previous definitions concern average properties of the primes
regarded as elements of  infinite progressions. However, despite the fact that the prime numbers are 
random objects subjected to average laws,  there exists  correlations between them. In particular, the pairwise
distribution of primes  is described by the function $\pi_2(k, x)$, that  counts 
the number of primes $p \leq x$ such that $p+ k$ is also a prime 

\beq
\pi_{2}(k,x) = \# \{ p, \, p \leq x , \,    p, p+k: {\rm primes} \}.
\label{321}
\eeq
The case $k=2$ gives the counting function of twin primes. 
Hardy and Littlewood conjectured the asymptotic behavior of  $\pi_2(k, x)$ \cite{HL}
\beq
\pi_2(k,x) \sim  C(k) Li_2(x)  \rightarrow   C(k)  \frac{ x}{ \log^2 \, x}, \quad x \rightarrow \infty, 
\label{34}
\eeq
where $Li_2(x) = \int_2^x dt/\log^2 t$, and $C(k)$ are the Hardy-Littlewood constants
\beq
C(k) = 0 \;  \; (k :  { \rm odd}),   \quad  
C(k) = C_2  \prod_{p>2, p |k} \left( \frac{ p-1}{p-2} \right)  \;  \;  (k :  {\rm even}), 
\label{35}
\eeq
with
\beq
C_2 = C(2) =  2 \prod_{q>2} \left( 1- \frac{ 1}{ (q-1)^2} \right) = 2 \times 0.6601618158..
\label{ctwin}
\eeq
The $q$-product in Eq.(\ref{ctwin})  runs  over all odd primes  and  the $p$-product  in Eq. (\ref{35}) runs  over all prime divisors
of $k$ except $2$. $C_2/2$ is called   the twin prime constant and has been computed with more
than 40 decimals. A beautiful heuristic derivation  of this result was found by Keating \cite{K93}, who
also derived the  asymptotic behavior of the function $C(k)$,  that in turn was important
to obtain  the GUE statistics of the Riemann zeros from the theory of Quantum Chaos (see Appendix A3). 

In the study of  entanglement of the Prime state, we have to deal with 
pairwise  correlations among arithmetic progressions, that leads us to 
another definition.  Let us consider
two arithmetic progressions  $a n + b$ and $a n'  + b'$, where $a$ is coprime to $b$ and $b'$
($b$ and $b'$ does not have to be coprimes among them). We shall define $\pi_{a; b, b'}(x)$ 
as  the number of prime pairs of the form $a n + b$ and $a n + b'$ (same $n$),  which are less or equal to $x$
(see  \cite{BK13}  for similar  functions)
\beq
\pi_{a;  b, b'}(x) = \# \{(p, p'), \,   p= a n+b,  p'= a n + b', \; p, p' : {\rm primes} , \; \; p, p' \leq x,  \; \; b \neq b',  \; \;  \; gcd(a,b) = gcd(a, b')  =1 \} .
\label{pab}
\eeq
The  pairs $(p, p')$  differ in $|b-b'|$, but not all the pairs that differ in that quantity
contribute to $\pi_{a;  b, b'}(x)$, because they may not be of the form  $(p= a n+b,  p'= a n + b')$.
In appendix A1, we give  an example. We have found numerically the following asymptotic behavior of 
$\pi_{a, b, b'}(x)$ (see Appendix A2)

\beq
\pi_{a; b,b'  }(x) \rightarrow \frac{C(|b - b'|) }{\phi(a) }   Li_2(x), \qquad x \rightarrow \infty, 
\label{36}
\eeq
that is a combination of Eqs.(\ref{32}) and (\ref{34}).  Indeed, the denominator $\phi(a)$
comes from the fact that one is  dealing with arithmetic series modulo $a$, while the remaining term
comes from having two  primes separated by $k = |b - b'|$.  Note that $k$ is always even
because $b, b'$ are both odd.

\subsection{The density matrix $\rho_A$ }

Equipped with the previous definitions we now embark into  the task of expressing 
the density matrix $\rho_A$ (Eq. (\ref{6})) in terms of the number theoretical functions
$\chi_\pi(a),  \pi(x), \pi_{a,b}(x), \pi_{a;b,b'}(x)$. The derivation is a bit technical 
and is left  to  Appendix A1. Here we state the main results.  $\rho_A$
is a matrix of dimension $M= 2^m$, where $m$ is the number of qubits
in $A$. The indices of $\rho^A_{a,a'}$  are  given 
an even/odd order, that is $a = (0,2,\dots, 2^{m}-2; 1,3, \dots, 2^m-1)$,
so that $\rho_A$ has the  block structure

\beq
\rho^A_{a, a'}  =  \frac{1}{ \pi(N)} 
\left( 
\begin{array}{ccc}
\delta_{a,2} \delta_{a', 2} &  & \delta_{a,2}  \, \chi_{\pi}(a') \\ 
\delta_{a',2} \,  \chi_{\pi}(a) &  & \delta_{a, a'} \pi_{M,a}(N) + ( 1 - \delta_{a, a'}) \pi_{M;  a, a'}( N) \\
\end{array}
\right). 
\label{22a}
\eeq
One can verify that 

\beq
{\rm tr}  \, \rho_A = \frac{1}{ \pi(N)}  \left(  1+ \sum_{a: {\rm odd} } \pi_{M, a}(N) \right)  = 1, 
\label{221}
\eeq
which follows from  the fact that  prime numbers less or equal to  $N$ 
are either 2, or of the form $a + M b$, with $a$ an odd number less than $M$. 
Eq. (\ref{22a}) shows that $\rho^A_{a, a'}$ vanishes if 
$a$ and $a'$ are even numbers different from 2. We shall then define
a truncated matrix $\hat{\rho}_A$ of dimension $\hat{M} = 2^{m-1}+ 1$,
which carries the same information as $\rho_A$, and whose 
labels  are $a=(2, {\rm odd})$, with the odd values in the interval $[1, 2^m-1]$, 

\beq
\hat{\rho}^A_{a, a'}  =  \frac{1}{ \pi(N)} 
\left( 
\begin{array}{cc}
1 &  \chi_{\pi}(a') \\ 
 \chi_{\pi}(a) & \delta_{a, a'} \pi_{M,a}(N) + ( 1 - \delta_{a, a'}) 
\pi_{M;  a, a'}( N) \\
\end{array}
\right).
\label{27b}
\eeq
In the limit $N \gg 1$  the contribution of $a=2$ to the entanglement
is negligible, so we can drop it and restrict ourselves to the density matrix involving
the odd values of $a= 1,3, \dots , M-1$

 \beq
{\rho}_{A; a , a'}^{\rm odd}  \equiv   \frac{1}{ \pi(N)-1} 
(  \delta_{a, a'} \pi_{M,a}(N) + ( 1 - \delta_{a, a'}) 
\pi_{M;  a, a'}( N) ), \qquad a, a' = 1, 3, \dots, M-1, 
\label{42b}
\eeq
that  satisfies the normalization condition ${\rm tr} \, {\rho}_{A}^{\rm odd} =1$.
In fact, Eq.(\ref{42b}) is the density matrix of a state formed by
the superposition of odd prime numbers, that is,  the odd prime state (note 
the  change of the denominator $\pi(N)$ in Eq.(\ref{27b}), for  $\pi(N)-1$ in Eq.(\ref{42b}), 
that does not modify the large asymptotic behavior of $\hat{\rho}_A$). 
In the limit $N \gg  \infty$, Eqs.(\ref{PNT}), (\ref{32}) and (\ref{36}) imply 

\beq
\pi(N) \rightarrow Li(N), \quad 
\pi_{M,a} \rightarrow \frac{1}{\phi(M)} Li(N), \quad \pi_{M; a, a'} \rightarrow  \frac{C(|a- a'|)}{\phi(M)} Li_2(N), \qquad N \rightarrow \infty,
\label{421}
\eeq
hence 

 \beq
\tilde{\rho}_{A; a , a'}^{\rm odd}  \rightarrow    \frac{1}{ \phi(M)} 
\left(  \delta_{a, a'} + ( 1 - \delta_{a, a'})   \frac{Li_2(N)}{ Li(N)}  C(|a-a'|)  \right), \qquad a, a' = 1, 3, \dots, M-1.
\label{43}
\eeq
The coprime divisors of $M= 2^m$ are the odd numbers $a=1,3, \dots, M-1$, so   $\phi(M) = 2^{m-1}$. 
To simplify the notations we shall write the  labels of $\tilde{\rho}_{A; a , a'}^{\rm odd}  $ as 

\beq
 a = 2 i - 1, \quad i=1, 2, \dots,  d = 2^{m-1} 
\label{44}
\eeq
and define

\beq
\ell_N  \equiv \frac{Li_2(N)}{ Li(N)} \rightarrow \frac{1}{ \log N} = \frac{1}{ n \log 2}, \quad N \rightarrow \infty.
\label{45}
\eeq
Eq. (\ref{43}) implies that $\tilde{\rho}_{A; a , a'}^{\rm odd}  $ approaches asymptotically the matrix 

 \beq
\bar{\rho}_{A; i , j}  \equiv   \frac{1}{d} 
\left(  \delta_{i, j} + ( 1 - \delta_{i, j})   \ell_N   C(2|i-j|)  \right), \qquad i, j = 1, 2, \dots, d, 
\label{46}
\eeq
whose diagonal entries are  equal to $1/d$ and off diagonal entries are given by
the Hardy-Littlewood function $C(k)$ (Eq.(\ref{35})). Therefore,  
in the asymptotic limit $N \rightarrow \infty$,  the bipartite entanglement of the prime state 
is encoded into the pairwise distribution  of prime numbers. Indeed, suppose
that the prime numbers  were pairwise uncorrelated, which amounts to the equation  $C(k) = 1$ (see Appendix A3). 
In this were so,  the  eigenvalues of Eq. (\ref{46}) would consist of  a dominant value  
 $\lambda_1$,  and $d-1$ degenerated values $\lambda_i \;(i=2, \dots, d)$ (see Appendix A4), 

\beq
{\rm If} \;  \; C(k) = 1 \Longrightarrow \lambda_1 = \frac{1 + \ell_N  (d-1)}{d}, \qquad \lambda_i = \frac{ 1 - \ell_N }{d} \; (i=2, \dots, d).
\label{461}
\eeq
In the limit $d \gg 1$, i.e. $m\gg 1$, the Renyi entropies are controlled by the largest eigenvalue $\lambda_1$,

\barray
{\rm tr} \, \bar{\rho}_A^\alpha & \simeq  & \ell^\alpha_N  \rightarrow \frac{1}{( n \log2)^\alpha},  
\, \quad S^{(\alpha)} \simeq \frac{\alpha}{1 - \alpha} \log \ell_N  \simeq
 \frac{\alpha}{\alpha -1} \log ( n \log 2), 
\label{462}
\earray   
while the von Neumann entropy is controlled by the degenerate eigenvalues 

\beq
S_A \simeq ( 1 - \ell_N  ) \log d \rightarrow  \left( 1 - \frac{1}{ n \log 2} \right)  (m -1)  \log 2 + \dots 
\label{463}
\eeq
that in the large $n$ limit converges towards the highest   entropy 
of a system with $d$  states,  that is $\log d$. Hence, the absence of correlations
between the primes would yield a density matrix with no information at all, that is
purely random \cite{P93}. However the primes are pairwise correlated, and  the amount
of correlation can be measured by the von Neumann and Renyi entropies of the
density matrix in Eq. (\ref{46}). More concretely, let us write Eq. (\ref{46}) as

\beq
\bar{\rho}_A = \frac{1}{d} ( {\bf 1} + \ell_N {\bf C}_m )
\label{464}
\eeq
where ${\bf C}_m$ is the Toeplitz matrix constructed with  $C(2 k)$, 

\beq
({\bf C}_m)_{i,j} = ( 1 - \delta_{i,j} ) C( 2 |i-j|), \quad i, j =1, 2, \dots, d=2^{m-1}.
\label{465}
\eeq
The Renyi entropies in Eq. (\ref{8}), associated to Eq. (\ref{464})  can then be computed 
from the traces of the powers of ${\bf C}_m$,

\beq
{\rm Tr} \,  \bar{\rho}_A^\alpha  = \frac{1}{d^\alpha } \left( d + \sum_{r=2}^\alpha 
\left( \begin{array}{c} 
\alpha \\
r \\
\end{array} \right) 
\ell_N^r   \, {\rm Tr} \,  {\bf C}_m^r \right) 
\label{466}
\eeq
where we use that ${\bf C}_m$ is traceless.  Eq. (\ref{464}) can be used  to express
${\bf C}_m$ in terms of $\bar{\rho}_A$ and the identity matrix, and then 
${\rm Tr} \,  {\bf C}_m^r$ can be obtained as a combination of  ${\rm Tr} \,  \bar{\rho}_A^\alpha$,
with $\alpha \leq r$.  The computation of the entanglement entropy $S_A$ is rather  complicate 
since it requires the knowledge of the eigenvalues of $ {\bf C}_m$. 

\subsection{Renyi-2 entropy/Purity}

Let us consider the Renyi entropy $S^{(2)}_A$,  or equivalently the purity
${\rm Tr} \, \rho_A^2$. The exact expression can be obtained from Eq. (\ref{22a}) or Eq. (\ref{27b}), 

\ba
P_A \equiv  {\rm Tr}_A \, {\rho}_A^2   & = &      \frac{1}{ \pi^2(N) } 
\left(  2 \pi(M) - 1  + \sum_a  \pi^2_{M,a}(N) +  \sum_{a \neq a'} 
\pi^2_{M;  a, a'}( N)  \right).  
\label{30}
\ea
while  for the density matrix (\ref{464}) one has 

\beq
\bar{P}_A \equiv {\rm Tr}_A \, \bar{\rho}_A^2    =       \frac{1}{d^2} 
\left( d + \ell_N^2 \, \sum_{i \neq j}^d  C^2(2 |i -j|) \right). 
\label{37}
\eeq
The asymptotic behavior of the sum is derived in Appendix A3 (see   Eq.(\ref{k37})), 

\beq
\sum_{i \neq j}^d  C^2(2 |i -j|)   \simeq \frac{2 \alpha^2}{\alpha_2} d^2, \qquad d \rightarrow \infty, 
\label{371}
\eeq
where 

\beq
\frac{2 \alpha^2}{\alpha_2} =    4  \prod_{p>2} \left( 1 + \frac{1}{(p-1)^3}  \right)   = 4.60192...
\label{391}
\eeq
which leads to

 \ba
 {\rm Tr}_A \, \bar{\rho}_A^2   &  \sim &   
 \frac{ 2 \alpha^2/\alpha_2}{(n \log 2 )^2},    \qquad  n= 2m  \gg 1,   
\label{39}
\ea
If the prime numbers where uncorrelated, 
the constant in the numerator of Eq. (\ref{39})  would be  equal to 1 (take $\alpha =2$ in   Eq.(\ref{462})). 
That this is not the case is a signature of correlations in the primes. 
The appearance of the denominator $(p-1)^3$ in Eq.(\ref{391})  
is related to  the fact that in a bipartition  $A/B$,  
one has to consider triplets of primes, say $p, q, r$, that are split  into  
$p q$ and $p r$ (see Appendix A3). This fact  suggests that Eq. (\ref{391}) is  another 
 measure of correlation among prime number analogue to the twin prime constant $C_2/2$.

\begin{figure}[h]
\centering
\includegraphics[width=0.35\textwidth]{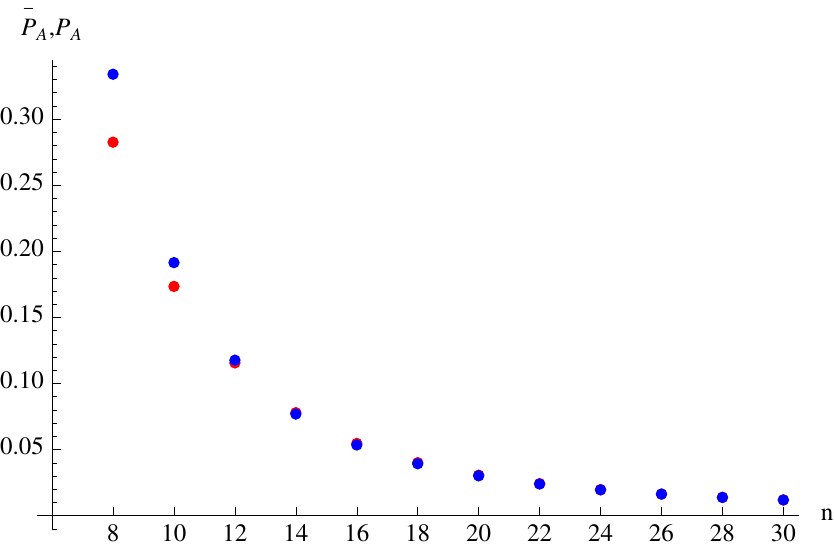}
\hspace{0.3cm}
\includegraphics[width=0.35\textwidth]{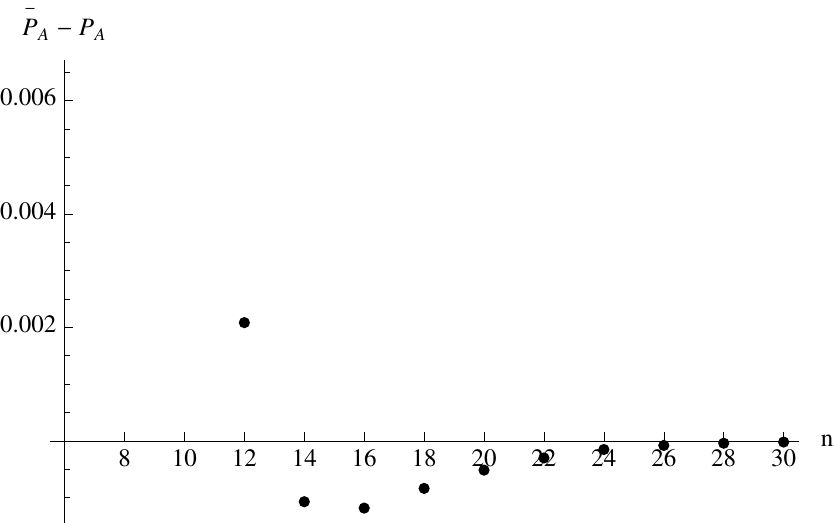}
\caption{Purity as a measure of entanglement for the Prime state. Left: Purity  for $n=8-30$ computed with the exact density matrix $\rho_A$  in Eq. (\ref{22a}) 
 (red on line) and the density matrix $\bar{\rho}_A$ in Eq. (\ref{464}) (blue of line). Right: difference
$\bar{P}_A - P_{A}$. 
 } 
\label{purity} 
\end{figure}

Fig. \ref{purity}  shows  the numerical results for  $P_A, \bar{P}_A$ and $P_A - \bar{P}_A$ 
for an equal bipartition with  $n = 2m$ with $n= 8$ up to to 30. Notice that the 
$P_A -\bar{P}_A$ decreases for increasing values of $n$, that confirms
the accuracy  of the approximation of $\rho_A$ by $\bar{\rho}_A$.

\subsection{ Numerical results for the von Neumann entanglement entropy}

The von Neumann entropy provides a figure of merit for quantifying
entanglement which has extensive  used in other quantum contexts, as 
in the fields of Condensed Matter or of black hole physics. 
Quite remarkably, it has proven to be an adequate instrument to
analyze the entanglement of the ground state of spin Hamiltonians and
it is used to detect and classify quantum phase transitions. In such situations, the scaling of the von Neumann entropy 
with the size of the system obeys the so-called "area law". 
A similar behavior is encountered in particle physics and in the holographic description of black holes.
Furthermore, the von Neumann entropy has an operational meaning. It describes
the number of singlets that can be distilled from a given state as the available
copies of that state go to infinity. It is, thus, natural to investigate
how much entropy of entanglement is carried by the Prime state.

From a technical point of view, the von Neumann entropy is difficult to calculate. 
The fact that we need to compute a logarithm of the reduced density matrix
makes an analytical approach not at all trivial. Here, we shall present numerical
results on this entanglement entropy for the Prime state up to $n=30$. To be precise,
we consider the Prime state at different values of $n$, $n=4, \ldots, 30$
and shall take its bi-partitions. Given a bi-partition, we compute the reduced 
density matrix for such a partition and then compute its associated
von Neumann entropy.

\begin{figure}[h]
\centering
\vspace{0.6 cm}
\includegraphics[width=0.35\textwidth]{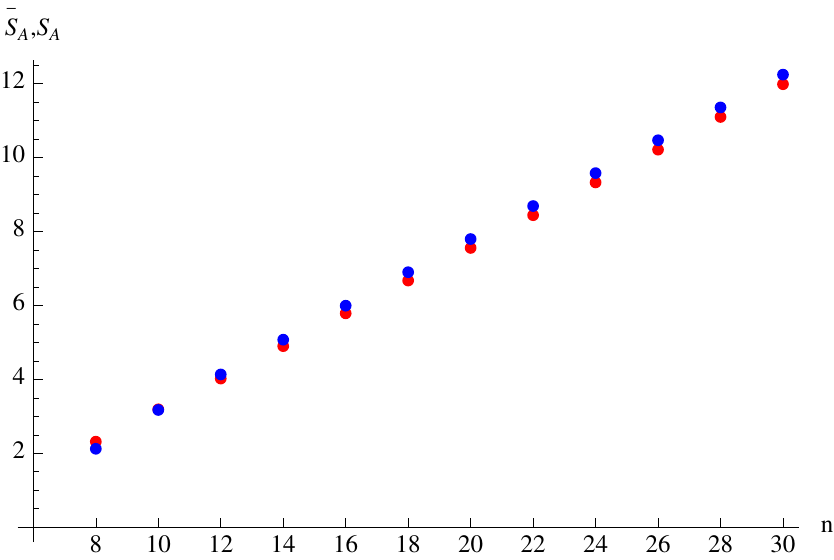}
\hspace{0.3cm}
\includegraphics[width=0.35\textwidth]{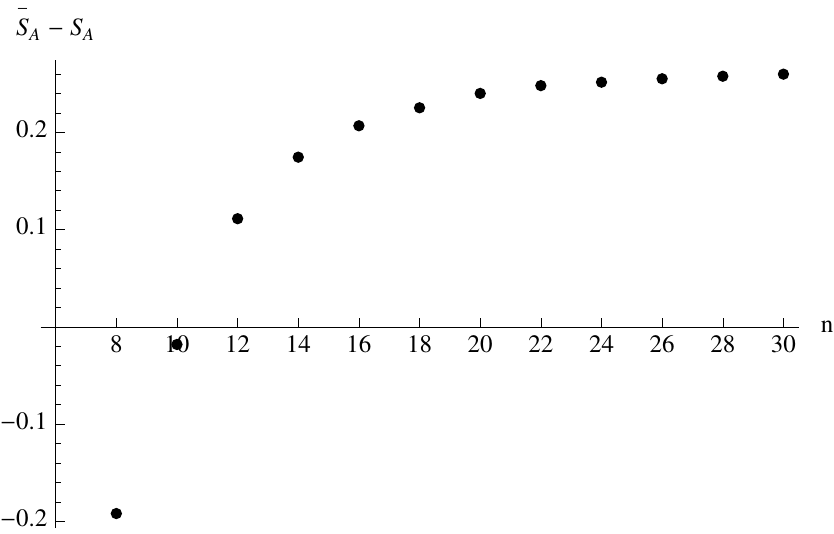}
\caption{Von Neumann entropy as a measure of entanglement for the Prime state. Left: Entanglement entropy  for $n=8-30$ computed with the exact density matrix $\rho_A$
 (red on line) and the approximation  Eq. (\ref{464}) (blue of line). Right: differences
$S_A - \bar{S}_{A}$. 
 } 
\label{von-Neumann} 
\end{figure}

Our main numerical result on the von Neumann entropy of the Prime state $|\Pmath_n \rangle$
is the scaling it shows as $n$ increases. This is shown in Fig. \ref{von-Neumann},
where we plot the actual results and a best fit to a line, $S(n)=\alpha n/2 + \beta$. We find that the
slope of this best fit is $\alpha=.886(1)$. Note that the absolute maximal entanglement
would correspond to the maximum possible slope $\alpha=1$. It follows that the Prime state seems to present
maximal scaling, which is proportional to $n$, with a coefficient which is less than 1. 
As in the case of the Renyi -2 entropy, we interpret this result as being due to
the existence of correlations in the primes. In section III C we obtain an analytic estimate
of the constant $\alpha$, based on the entanglement spectrum of the density matrix $\bar{\rho}_A$.


Let us emphasize that it is expected that the Prime state should carry very  large
entanglement. If this were not the case, we could apply  standard efficient techniques like
Matrix Product States to handle its properties (for a review see \cite{CV}). The fact that the Prime
state is so highly entangled is just preventing the applicability of some powerful tools  
which are useful on standard physical systems. 
To have a better understanding of this point, we compare in Table \ref{scalings} 
the known scaling of entanglement in different systems.

\begin{table}
\begin{tabular}{|c|c|}
\hline
infinite non-critical spin chains & $S\sim \frac{c}{6} \log \xi$, $\xi$  correlation length \\
\hline
critical spin chains, block of size $L$ & $S\sim \frac{c}{3} \log \L$ \\
\hline
2D systems, square of size $L$ & $S\sim L$\\
\hline
Prime state of $n$-bits & $S\sim \alpha n/2$, $\alpha\sim .886(1)$\\
\hline
maximally entangled states & $S\sim n/2$\\
\hline
\end{tabular}
\caption{Scaling of von Neumann entropy for different quantum systems. In
the case of spin chains, $c$ is the central charge of the underlying CFT.}
\label{scalings}
\end{table}

\begin{figure}[h]
\centering
\vspace{0.6 cm}
\includegraphics[width=0.35\textwidth]{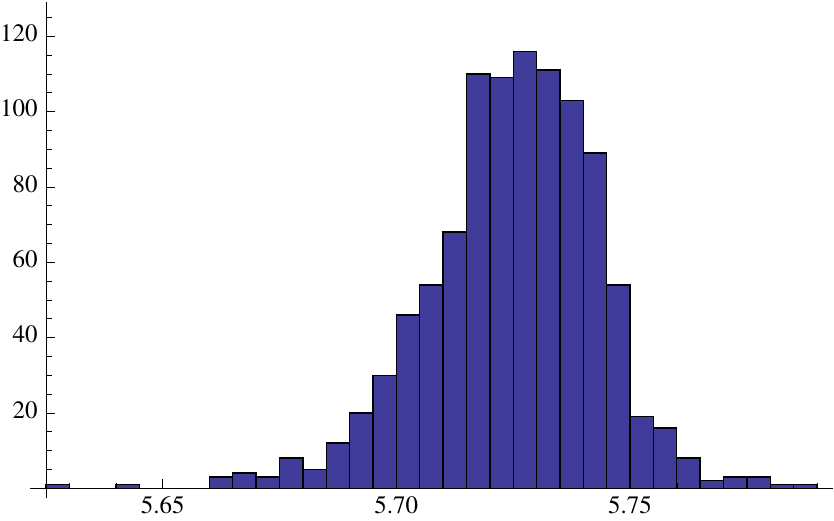}
\caption{Histogram showing the entanglement entropy of 1000 random partitions of the Prime state with $n=20$ quits
(the abscissa corresponds to  the entropy). 
The right-most point, with largest entropy, corresponds to the natural partition.}
\label{entanglement-histogram} 
\end{figure}

The interpretations of large classical and quantum entropies are related to {\em surprise}. We may rephrase
the maximal scaling of entanglement in the following way.
Suppose that we know the first
$n/2$ digits of a prime. Then, knowing the second $n/2$ digits provides
a classical surprise which is almost maximal. Then, let us think of the quantum superposition of primes. The correlation between all the first $n/2$ digits
and the rest is almost maximal. We cannot disentangle the first and second
part of the register.
 
Let us now turn to  the study of different partitions of the Prime state.
We consider random bi-partitions of the Prime state which need not
be sequential, that is the party $A$ is made out of $n/2$
qubits, randomly chosen out of the $n$ available ones. 
Fig.\ref{entanglement-histogram}  shows the von Neumann entropy for  large sets
of  bi-partitions of  $n=20$ qubits.
 It is  remarkable that the maximum entropy is found for the natural partition of the Prime state.
This fact is systematically observed for all values of $n$  we have been able to checked. 
It is unclear if this is a fundamental fact about the distribution of primes. 
A similar result is obtained if purity is analyzed. Again the natural partition shows more entanglement than any other one.

\subsection{Majorization of twin primes by primes}

Twin primes form a series which has been thoroughly analyzed in 
Number Theory. It was shown in section  II that a 
 Twin Prime quantum state can be constructed using the
 same technology as for the Prime state. We may as well
 ask whether the Twin Prime state is more or less entangled
 than the Prime state. The answer to this question
 is shown in Table \ref{comparisonprimetwinprime}.
 
  \begin{table}
\begin{tabular}{|c|c|c|}
\hline
n&S(prime)& S(twin prime) \\
\hline
10& 3.1900& 3.3450\\
12& 4.0220& 4.5221\\
14& 4.8993& 5.4438\\
16& 5.7872& 6.4812\\
18& 6.6748& 7.4908\\
20& 7.5574& 8.4834\\
22& 8.4428& 9.4796\\
24& 9.3301& 10.4729\\
26& 10.2159& 11.4644\\
28& 11.1018& 12.4551\\
30& 11.9876&\\
\hline
\end{tabular}
\caption{Comparison between the  von Neumann entropy for the Prime state and 
the Twin Prime state. The Twin Prime state carries more correlation than primes.}
\label{comparisonprimetwinprime}
\end{table}

The Twin Prime state appears to carry more entanglement than the
Prime state. This is somewhat reasonable. Twin primes are more scarce than primes, $N/\log^ 2 N$ as compare to $N/\log N$. Yet, the maximum of entropy requires only $N^{1/2}$ superpositions, which is far less than the actual
number of twin primes. It seems that the more dilute twin primes are then enough to carry more surprise in their appearances. 

We may now wonder whether the probability density described by the entanglement
spectrum of the Prime state, $\lambda_i^{(p)}$ with $i=1,\ldots, 2^ {n/2}$, is strongly ordered with respect to the
equivalent entanglement spectrum in the Twin Prime state,
$ \lambda_i^{(p,p+2)}$ with $i=1,\ldots,2^ {n/2}$. This is indeed the case. It is easy to verify that
for large $n$ the following $2^ {n/2}$ relations hold
\begin{eqnarray}
\nonumber
\lambda^{(p)}_1 & > & \lambda^{(p,p+2)}_1\\
\nonumber
\lambda^{(p)}_1+\lambda^{(p)}_2 & > & \lambda^{(p,p+2)_1}+\lambda^{(p,p+2)}_2\\
\nonumber
\ldots &  \ldots\\
\lambda^{(p)}_1 +\ldots +\lambda^ {(p)}_{2^{n/2}}& > &
\lambda^{(p,p+2)}_1 +\ldots +\lambda^ {(p,p+2)}_{2^{n/2}} .
\end{eqnarray}
Thus, the Prime state spectrum majorizes the Twin Prime spectrum
\be
 \vec\lambda^{(p,p+2)}\prec  \vec\lambda^{(p)}.
\ee
This is remarkable sense of order usually named majorization.
From an operational point of view, majorization is stated about the possibility of connecting quantum states using Local Operations and Classical Communication. In our case, the general theorems mean that if two parties  $A$ and $B$ can carry unitaries
on their local part of a quantum state and can also communicate classically, then the Twin Prime state can be transformed into the Prime state. This is
a consequence of the majorization relations that they obey.

We have further explored whether there are majorization relations for triples of primes, those with the form $(p,p+2,p+6)$ or $(p,p+4,p+6)$. We have analyzed
the first of these cases and found again a relation of majorization
\be
\vec \lambda^{(p,p+2,p+6)} \prec
\vec \lambda^{(p,p+2)} \prec
\vec \lambda^{(p)}.
\ee
if $n$ is sufficiently large. At small $n$ the fluctuations of primes
do spoil the majorization relation, but this transient situation goes
away monotonically as $n$ grows.

It is tantalizing to conjecture that an infinite series of majorization
relations, that is a deep sense of ordering, settles in the set 
of subseries of primes. The more sparse sub-series of primes will typically majorize
the denser ones.

\section{Analytical approximation to the entanglement spectrum}

The different measures of entanglement we have computed for the Prime state
are ultimately based on the
precise distribution of eigenvalues of its Schmidt decomposition.
As a matter of fact, the spacing and degeneracy of these eigenvalues contain
deep information on the quantum correlations of the state. We shall shortly show
that the detailed analysis of this spectrum of eigenvalues hints at an analytical approximation to the entanglement properties of the Prime state, that we shall later exploit.

The entanglement spectrum of the density matrix $\rho_A$ is defined in terms of its eigenvalues $\lambda_i$
as follows
\beq
\varepsilon_i = - \log \lambda_i, \qquad i=1,  \ldots, 2^m,
\label{ee}
\eeq
and  they are the eigenvalues of the entanglement Hamiltonian $H_E$
defined as $\rho_A = e^{ - H_E}$. The states with lowest entanglement energy are 
therefore  those with highest values of $\lambda_i$. This definition was introduced in the context
of the Fractional Quantum Hall effect, where the low energy entanglement spectrum of the ground state, 
at several filling fractions, was shown to  correspond to the Conformal Field Theory that describes 
the physical edge excitations of the system \cite{LH}. This result suggests  a  bulk-edge correspondence
in strongly correlated many-body systems that is reminiscent to  the holographic principle in String Theory. 


\begin{figure}[h]
\centering
\includegraphics[width=0.3\textwidth]{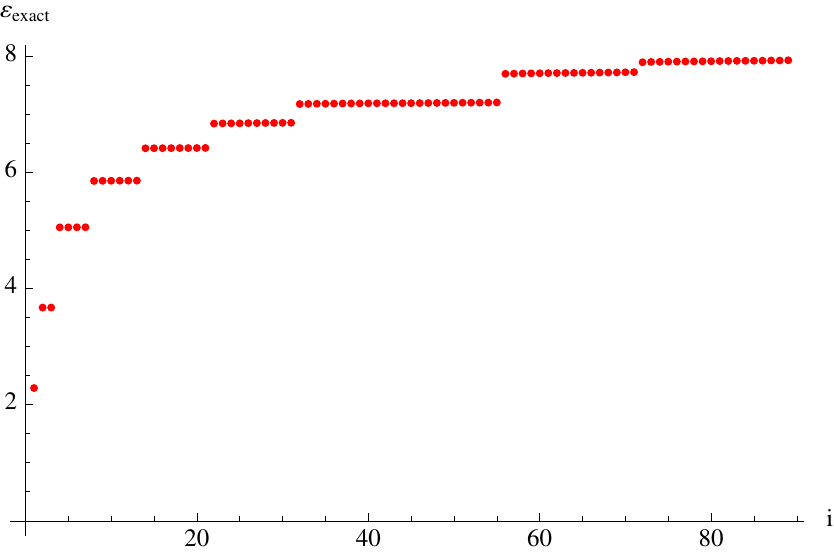} 
\hspace{0.5cm}
\includegraphics[width=0.30\textwidth]{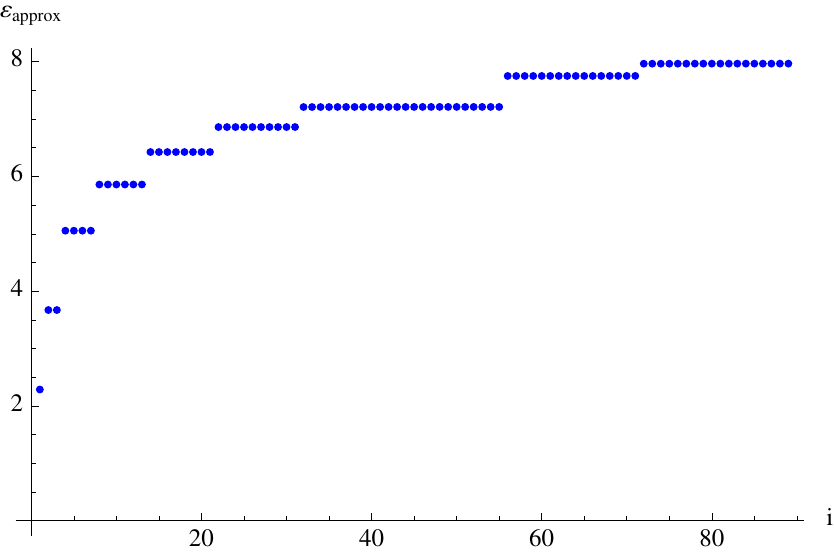}
\hspace{0.5cm}
\includegraphics[width=0.30\textwidth]{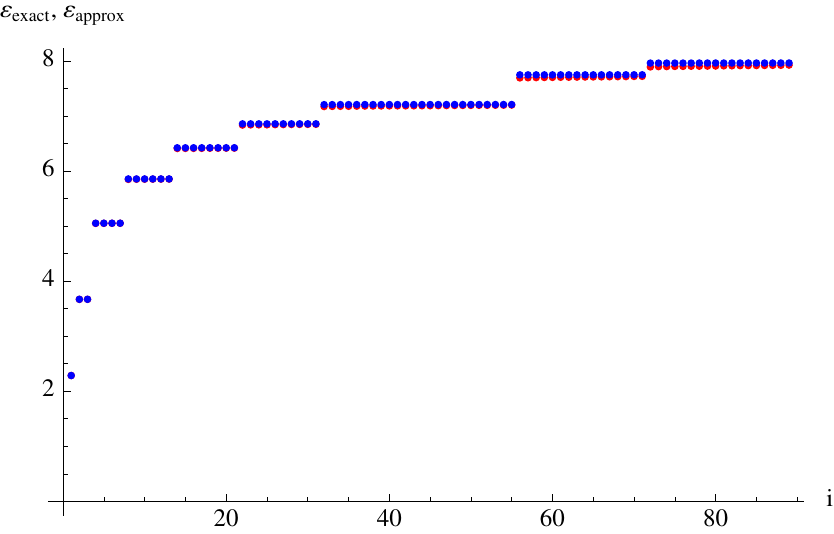}
\caption{Entanglement spectrum for $n=30$. It is plotted the lowest 89 states.
Left:  EE's of  $\rho_A$ (red on line). Right: EE's of  $\bar{\rho}_A$ (blue on line). Right: 
Superposition of EE of  $\rho_A$ and $\bar{\rho}_A$. 
 } 
\label{gaplot}  
\end{figure}

 We have computed  the entanglement spectrum  of the  Prime state using the exact density matrix $\rho_A$  of Eq. (\ref{22a}) 
 and the approximation $\bar{\rho}_A$ of Eq. 
 (\ref{46}). Fig. \ref{gaplot} shows the lowest 89 states for  $n=30$ qubits with a bipartition  $m =n/2=15$. We first notice
 that the exact and the approximated values of $\varepsilon_i$ agree rather well, specially in the low energy part, and
 that the agreement improves increasing the number of qubits $n$. The more interesting feature is the appearance
 of nearly degenerated energy eigenstates,  particularly visible for $\bar{\rho}_A$. We shall now obtain an analytical approximation to these
 eigenvalues and their degeneracies. 
 
\begin{figure}[h]
\centering
\vspace{0.6 cm}
\includegraphics[width=0.4\textwidth]{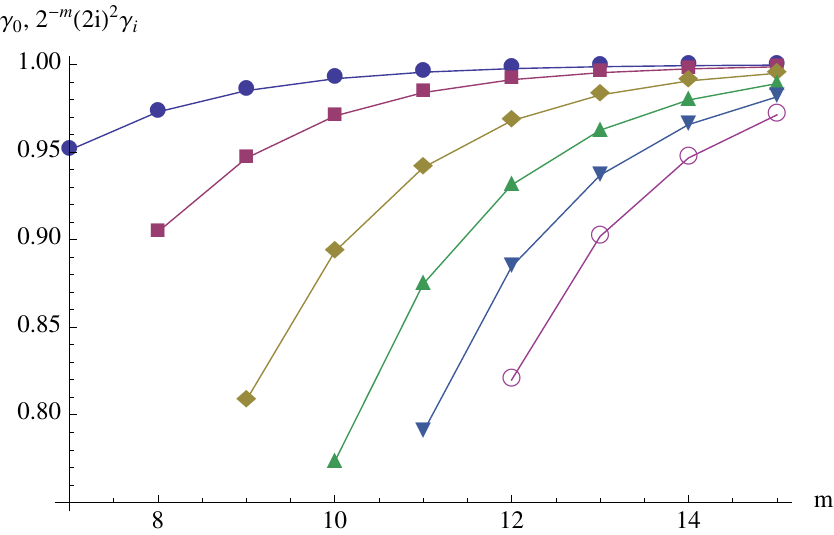}
\caption{Verification of the relation (\ref{gama1}) for the eigenvalues of the matrix ${\bf C}_m$
as a function of $m=7, \ldots, 15$. From top to bottom we plot $\gamma_0, \gamma_1, \dots, \gamma_5$. 
Notice that the relation becomes exact  in the limit $m \rightarrow \infty$.}
\label{gamma} 
\end{figure}

 Note that the relation in Eq. (\ref{464}) implies
 that the  degeneracy in the entanglement spectrum must  be originated in the eigenvalues $\gamma_i$ of the  matrix ${\bf C}$. In our case, we work with a finite amount of primes, so that
 we have  ${\bf C}_m$, where $m=n/2$, that approaches ${\bf C}$ as the number
 of qubits in the Prime state goes to infinity.  
 The lowest eigenvalue, $\gamma_0$, is unique but the  positive higher eigenvalues $\gamma_i \; (i \geq 1)$
 have a degeneracy $2i$ and are  given by 
\beq
\gamma_0  =  2^m,   \qquad 
\gamma_i= \frac{2^m}{( 2 i)^2} \quad( i=1,2, \dots, k_m),   \label{gama1}
\eeq
where the value of  $k_m$ associated to  the largest eigenvalue will be computed shortly.
Fig. \ref{gamma} shows a numerical check of this relation as a function of the number of qubits $m$ of the block $A$. 
Eqs.(\ref{464}) and (\ref{gama1}), imply the following approximation for the eigenvalues of the density matrix $\bar{\rho}_A$, 
\beq
\lambda_0  =  2^{1-m} ( 1 + \ell_N 2^m),   \qquad 
\lambda_i= 2^{1-m} \left( 1+  \ell_N \frac{2^m}{( 2 i)^2} \right) \quad( i=1,2, \dots, k_m),    \label{gama2}
\eeq
where $\ell_N$ is given by eq.(\ref{45}) with  $n = 2 m$.  To fix the value of $k_m$ we shall impose
the normalization of the density matrix $\bar{\rho}_A$, that is
\beq
\lambda_0 + \sum_{i=1}^{k_m} 2 i \, \lambda_i =1, 
\label{gama3}
\eeq
 which for large values of $m$ yields
\beq
k_m \simeq \kappa_0   \, 2^{m/2} ( 1 - \beta/m), \quad \kappa_0= \sqrt{ \frac{3}{8} }, \quad \kappa_1 = 1.0035..
\label{gama4}
\eeq 
The matrix ${\bf C}_m$ is traceless, so there are   negative eigenvalues whose sum compensates the
sum of the positive eigenvalues in Eq. (\ref{gama1}). The magnitude of the former ones   is small  
which implies that the corresponding  eigenvalues of $\bar{\rho}_A$ will also be small. 
We shall show below  that the positive eigenvalues (\ref{gama1})  dominate the traces of the powers of the matrix ${\bf C}_m$, 
and consequently the  Renyi and von Neumann entropies. Let us first compute 
\barray 
{\rm  Tr} \, {\bf C}_m^s  & \approx &  \gamma_0^s + \sum_{i =1}^{k_m} 2 i \, \gamma_i^s 
= 2^{m s} \left( 1 +  \sum_{i =1}^{k_m} \frac{1}{(2 i)^{2 s -1}}   \right)  
\xrightarrow{\scriptscriptstyle m\to\infty}
2^{m s} \left( 1 +   \frac{\zeta( 2 s -1)}{2 ^{2 s -1}}    \right), \quad
s > 1.
\label{gama5}
\earray
where we took the limit  $m \rightarrow \infty$ in the sum, which  gives rise to the Riemann zeta function $\zeta(2s-1)$. 
Notice that the series diverges if $s \rightarrow 1$ because we have dropped the negative eigenvalues.  For  $s=2$, one obtains
\barray 
{\rm  Tr} \, {\bf C}_m^2  \xrightarrow{\scriptscriptstyle m\to\infty}   2^{2 m} \left( 1 +   \frac{\zeta( 3)}{8}    \right) =  2^{2 m} \times  1.15026..
\label{gama6}
\earray
which agrees to order $10^{-4}$ with   the exact  asymptotic expression of Eq. (\ref{371}) derived in  Appendix A3
\be 
{\rm  Tr} \, {\bf C}_m^2   
\xrightarrow{\scriptscriptstyle m\to\infty}
 2^{ 2 m} \prod_{p>2} \left( 1 + \frac{1}{(p-1)^3} \right)  =    2^{2 m} \times  1.15048...
\label{gama7}
\ee
Based on this numerical result  we conjecture the following asymptotic formula for any $s \geq 2$,

\be
{\rm  Tr} \, {\bf C}_m^s 
\xrightarrow{\scriptscriptstyle m\to\infty} 2^{ m s} \prod_{p>2} \left( 1 + \frac{1}{(p-1)^{2 s-1}} \right).
\label{gama8}
\ee
The difference between (\ref{gama8}) and (\ref{gama5}) is 

\beq
\prod_{p>2} \left( 1 + \frac{1}{(p-1)^{2 s-1}} \right) =  \left( 1 +   \frac{\zeta( 2 s -1)}{2 ^{2 s -1}}    \right) + 
\frac{1}{12^{2 s -1}} - \frac{1}{14^{2 s -1}} + \dots, 
\label{gama9}
\eeq
that for $s=3$ explains the proximity  of the constants in Eqs. (\ref{gama6}) and (\ref{gama7}). Let us note that the zeta function term approximates 
exponentially well the infinite product over primes.
In table \ref{cmateig}  we compare Eqs. (\ref{gama8}) and (\ref{gama5}) with the numerical results
obtained by extrapolating  ${\rm  Tr} \, {\bf C}_m^s$ for $m=9, \dots, 15$.

 \begin{table}
\begin{tabular}{|c|c|c|c|}
\hline
$s$  & $ 1 +   \frac{\zeta( 2 s -1)}{2 ^{2 s -1}} $ &  $\prod_{p>2} (1 + (p-1)^{1- 2s})$ &  $\lim_{m \rightarrow \infty} 2^{-ms} {\rm  Tr} \, {\bf C}_d^2$  \\
\hline
2 & 1.15025711  & 1.15048076  & 1.15048162\\
3 & 1.03240399  & 1.03240618  & 1.03241334  \\
4 & 1.00787772  & 1.00787774 & 1.00788749 \\
5 & 1.00195704  & 1.00195704 &1.00197171  \\
\hline
\end{tabular}
\caption{Comparison of   $ \lim_{m \rightarrow \infty} 2^{-ms} {\rm  Tr} \, {\bf C}_m^2$ computed with Eqs. (\ref{gama5}), (\ref{gama8}) and
the numerical values obtained with the fit formula $a_0 + 2^{-m} \sum_{j=0}^3 a_j \, m^j $  with  $m=9, \dots, 15$.}
\label{cmateig}
\end{table}

To calculate the entanglement entropy we use Eq.(\ref{gama2}) and replace the sum over the eigenvalues by an integral

\barray
S_A  & \approx &  -  ( \lambda_0  \log \lambda_0 + \sum_{i=1}^{k_m} 2 i  \lambda_i  \log \lambda_i) 
\label{gama10} \\
& \approx & - \int_{1}^{k_m} dx \, x \;  2^{2-m} \left( 1 + \frac{ \ell_N 2^{m-2}}{x^2}  \right)  \log \left[ 2^{1-m}  \left( 1 + \frac{ \ell_N 2^{m-2}}{x^2}  \right) 
\right] .  
\nonumber  
\earray
Then  computing  the two leading terms in the asymptotic limit $m \rightarrow \infty$
we obtain an analytical approximation to the von Neumann entropy
\be
S_A \xrightarrow{\scriptscriptstyle m\to\infty}
 \frac{7}{8}  m + \frac{1}{4} \log m + {\rm cte},   \qquad m \rightarrow \infty.  
\ee 
For other values of the parameter $\kappa_0$,  giving $k_m$ in  Eq.(\ref{gama4}), the constant of the linear term in (\ref{gama10})
is given by $(1 + 16 \kappa_0^2)/8$, that becomes $7/8 = 0.875$  for $\kappa_0 = \sqrt{3/8}$. 
Observe that the latter  constant  is close to the numerical value $0.886(1)$ obtained from the spectrum of the
density matrix $\rho_A$ up to sizes $m=15  \; (n=30)$ given in Table \ref{scalings}.

\section{Novel number theoretical states}

The prime state can be generalized to many other states associated to
arithmetic functions. Consider for example the Moebius function $\mu(n)$.
Factorizing $n$ into products of primes, $n = p_1^{r_1} \dots p_s^{r_s}$, then $\mu(n)=0$  
if $n$ contains a prime to a power greater than one, $\exists \;  i,  r_i > 1$,  and if $n$
is the product of $s$ distinct primes, $r_i =1, \; \forall i$,  then  $ \mu(n)= (-1)^s$, 
\beq
n = p_1^{r_1} \dots p_s^{r_s}, \quad 
 \mu(n) = \left\{
 \begin{array}{cc}
 (-1)^{r_1 + \dots + r_s} & {\rm if} \; r_1 = \dots = r_s = 1, \\
 1 & n=1, \\
 0 & {\rm else} \\
 \end{array}
 \right. \ .
 \label{p1}
 \eeq
The Moebius state shall  be defined as 
\beq
|  \mu_n \rangle = \frac{1}{ \sqrt{C_{\mu, n}}}  \sum_{a < 2^n}  \mu(a) \, | a \rangle, \quad C_{\mu , n} = \sum_{a < 2^n} |\mu(a)|, 
\label{p2}
\eeq
where $C_{\mu, n}$ ensures  that $\langle \mu_n | \mu_n \rangle =1$. This constant
can be estimated using the quantum counting protocol employed  to compute  $\pi(2^N)$ \cite{LS}. 
The similarity between the prime state of Eq. (\ref{1}) and the Moebius  state of Eq. (\ref{p2}) is made manifest 
writing  Eq. (\ref{1}) in terms of  the characteristic function $\chi_\pi$
\beq
| \Pmath_n \rangle = \frac{ 1}{ \sqrt{\pi(2^n)} } \sum_{a   < 2^n}  \chi_\pi(a)  |a \rangle, \quad  \pi(2^n)  = \sum_{a < 2^n} \chi_\pi(a).
\label{p3}
\eeq
The Moebius state encodes quantum mechanically the information contained in $\mu(n)$.
As in the case of the Prime state one can extract  part of this information performing
measurements. An example of this method  is provided by the Mertens function \cite{Apostol} 
\beq
M(n) = \sum_{a \leq n} \mu(a), \quad n =1,2, \dots
\label{p4}
\eeq
 $M(2^n)$ can be measured
projecting the Moebius state into  the  Hadamard state of $n$ qubits
\beq
| H_n \rangle = \frac{1}{2^{n/2}} \sum_{a < 2^n} | a \rangle , 
\label{p5}
\eeq
that gives
\beq
\langle H_n | \mu_n \rangle = \frac{1}{ 2^{n/2}  C_n^{1/2}}   M(2^n). 
\label{p6}
\eeq
The constant $C_{\mu, n}$ is known from  quantum counting so $M(2^n)$ can be  determined
within some error inherent to  quantum measurements. The Riemann hypothesis
implies that $|M(x)| = O(x^{\frac{1}{2} + \epsilon}), \, \forall \epsilon >0$   \cite{Ti},
so one could try to verify its validity, as was explained in section II C using  the Prime state. 

In previous sections we have shown that the Prime state contains a high degree
of entanglement that is intimately connected to the pairwise  correlations between the prime numbers.
The Moebius state, and most of the  states built using arithmetic functions, are expected
to be entangled. One may ask  what is the relation between the arithmetic
functions and the measurements of entanglement of the corresponding states? 
We shall not consider here this problem, and only make some suggestions. 

The arithmetic quantum states of the form in Eqs. (\ref{1}) and (\ref{p2}) can
be generalized to the Dirichlet series, 

\beq
f(s) = \sum_{n=1}^\infty \frac{ f(n)}{ n^s}, \qquad \Re \, s > \sigma_c , 
\label{p7}
\eeq
where $f(n)$ is an arithmetic function,  and $s = \sigma + i t$ a complex variable whose real
part  $\sigma$ must be bigger than  some value $\sigma_c$ to ensure the convergence 
of the series \cite{Apostol}. An example of Eq. (\ref{p7}) is the Riemann zeta function 
$\zeta(s) = \sum_{n=1}^\infty 1/n^s$, where $\sigma_c = 1$.  We can associate to $f(s)$  the Dirichlet 
state 

\beq
|  f(s)_n \rangle = \frac{1}{ \sqrt{C_{f,n}(\sigma)}}  \sum_{a < 2^n}  \frac{ f(a)}{a^s}  \, | a \rangle, \quad C_{f, n}(\sigma)  = \sum_{a < 2^n} \frac{|f_n(a)|^2}{a^{2 \sigma}}, \quad
s = \sigma + i t.  \label{p8}
\eeq
In the limit $n \rightarrow \infty$, the normalization constant  $C_{f, n}(\sigma)$, becomes a Dirichlet
series whose radius of convergence will  be bigger than  $\sigma_c$. Consider again the case of $\zeta(s)$
where

\beq
\lim_{n \rightarrow \infty}  C_{\zeta, n}(\sigma)  = \sum_{n=1}^\infty  \frac{1}{n^{2 \sigma}} = \zeta(2 \sigma),  \qquad
\sigma  > \frac{1}{2}, 
\label{p9}
\eeq
that allows us to define a normalized state associated to $\zeta(s)$,   including  the critical  region $\frac{1}{2} <  \sigma \leq 1$, 
where Eq. (\ref{p7}) diverges and $\zeta(s)$  is defined by analytic extension. The Riemann state
can then be defined as

\beq
|  \zeta (s) \rangle =  \frac{1}{ \sqrt{\zeta(2 \sigma)}}  \sum_{n=1}^\infty   \frac{1}{n^s}  \, | n  \rangle, \quad  
\sigma > \frac{1}{2}. 
 \label{p10}
\eeq
We expect this state, or its finite qubit version in Eq. (\ref{p8}),  to exhibit different entanglement properties
in the critical region $\frac{1}{2} <  \sigma \leq 1$ and in  the non-critical one $\sigma >1$, that may in
turn be related to the properties of $\zeta(s)$ itself. 
Similarly,  in the limit $n \rightarrow \infty$ we can associate a quantum state to any Dirichlet
$L$-function with character $\chi$

\beq
L_\chi(s) = \sum_{n=1}^\infty \frac{\chi(n)}{ n^s}   \longrightarrow 
|  L _\chi (s) \rangle =  \frac{1}{ \sqrt{\bar{L}_\chi(2 \sigma)}}   \sum_{n=1}^\infty   \frac{\chi(n)}{n^s}  \, | n  \rangle, \quad  
\bar{L}_\chi(\sigma) = \sum_{n=1}^\infty \frac{ |\chi(n)|}{ n ^{2 \sigma}}. 
 \label{p11}
\eeq
These  examples shows that one can assign  entanglement properties
not only to prime numbers and arithmetic functions, but also   to the Riemann and
Dirichlet $L$- functions, opening a novel approach to Number Theory using the tools
of Quantum Information Theory.

\section{Conclusions}

Quantum Mechanics offers a completely new approach to Arithmetics, that
may  be called Quantum Arithmetics. 
The essential point is the possibility of producing a quantum
superposition of series of numbers, that can then be processed
and analyzed in parallel. In this paper, we have shown that the series of prime
numbers can be assigned a Prime state. This Prime state can
be created efficiently and can be exploited to find properties
of the series of prime numbers in a way that goes  beyond 
classical computational methods. As a relevant example, we have shown
that the Prime state can be used to verify the Riemann Hypothesis
in a more efficient way than any existing classical algorithm.

We have shown that the density matrix of the Prime state is
characterized, asymptotically,  by the Hardy-Littlewood constants
that parametrize  the pair correlations between prime numbers. 
This result establish a  link between between prime correlations
and entanglement of the Prime state. We have also studied
the figures of merit of entanglement in this state, such as 
purity,  von Neumann entropy and entanglement spectrum. 
The main result stays than
the Prime state entanglement is almost maximal. As
a consequence, it follows the inefficiency of using clever tensor 
networks algorithms to manipulate it.

\vspace{1 cm}

\indent \tn{Acknowledgements.} 
The authors are grateful to J. I. Cirac, A. C\'ordoba, 
S. Iblisdir and J. Keating for helpful comments and to A. Monr\`as for his help on the computation of entropies.
J. I. L. acknowledges  financial support  from FIS2011-16185, Grup de Recerca Consolidat 
ICREA-ACAD\`EMIA, and National Research Foundation \& Ministry of Education, Singapore; and G. S. 
from the grants FIS2012-33642, QUITEMAD and the Severo-Ochoa Program.


\section*{Appendix A1: Derivation of $\rho_A$}

In this section we shall find an explicit expression of the 
density matrix  $\rho_A$ in Eq. (\ref{6}) 
in terms of number theoretical functions whose asymptotic  properties
will be presented in Appendix A2. To do so we shall  
treat separately the blocks corresponding to  $m=1$ and $m >1$.
If $m=1$, the  matrix $\rho_A$ is  given by  \cite{LS}  

\beq
m=1: \;  \rho^{A}_{ 0,  0}  = \frac{1}{\pi(N)}, \quad   \rho^{A}_{ 1,  1}  = \frac{\pi(N) -1  }{\pi(N)},  \quad   \rho^{A}_{ 0,  1}  = \frac{1  }{\pi(N)}.
\label{10}
\eeq
These equations follow from the observation  that all the primes are odd numbers but  $p=2$,
which implies 

\ba
p= 2 & \rightarrow & \psi_{b,0} = \delta_{b, 1} , \label{11} \\
p > 2  & \rightarrow & \psi_{b,1}  = 
\left\{ 
\begin{array}{ll} 
1  &  {\rm iff} \; \;  p = 1 + 2 b : {\rm prime},  \\
0 & {\rm else}. \\
\end{array}
\right.  .
\nonumber
\ea
The numerators of $\rho^A_{a,a'}$,  
means  that 2 is a prime number ($(a, a')=(0,0)$), 
that there are $\pi(N)-1$ odd prime numbers ($(a, a')=(1,1)$)  and that $2$ and $3$ are primes ($(a, a')=(0,1)$). 

Let us next consider the cases where $m >1$.  Since $\psi_{a,b} =0$ or 1, $\forall a,b$, one can write
the diagonal entries of $\rho_A$ as 
\beq
\rho^A_{a, a}     =  \frac{1}{ \pi(N) } \sum_{b< 2^{m'}} \psi_{b,a}. 
\label{12}  
\eeq
In this equation the sum 
gives  the number of primes of the form $p=a + 2^m \, b$  less than $N$,
where $a \in [0, 2^m-1]$ is fixed  and $b$ varies  in the interval $[0, 2^{m'}-1]$. 
If $a$ is even so is $p$,  because $2^m b$ is always even or zero. Hence  $p$ is equal to 2,
and then $a=2$ and $b=0$, which yields

\beq
m> 1, \; a : {\rm even} \rightarrow \psi_{b,a}  = \delta_{b,0} \delta_{a,2}. 
\label{13}
\eeq
Note that the condition $m>1$ is required to allow $a$ to take the value 2. 
Using Eq.(\ref{13}) into Eq. (\ref{12}) one finds

\beq
m> 1, \; a : {\rm even} \rightarrow \rho^A_{a,a}  =  \frac{ \delta_{a,2}}{ \pi(N)}. 
\label{14}
\eeq

If $a$ is odd, the primes $p = a + 2^m b$, form an arithmetic progression
$c n + d \, (n=0,1, \dots)$ where $c = 2^m$ and $d= a$
are coprime numbers, that its with no common divisor.  Using the standard notation 
 $\pi_{c,d}(x)$ for the number of primes less or equal to $x$ contained  in the arithmetic
progression $c n + d$, where  $c$ and $d$ coprime numbers, we can then write

\beq
m> 1, \; a : {\rm odd} \rightarrow \rho^A_{a,a}  =  \frac{ \pi_{M,a}(N)}{ \pi(N)}, \quad M = 2^m.
\label{15}
\eeq
 Let us next consider the off-diagonal terms $\rho^A_{a, a'}$. 
 If $a \neq a'$ are both even, one can use Eq.(\ref{13}) to derive
 \beq
m> 1, \; a \neq  a'  : {\rm even} \rightarrow \rho^A_{a,a'}  =  0.
\label{16}
\eeq
If $a$ is even and $a'$ odd using again Eq.(\ref{13}) one finds

 \beq
m> 1, \; a : {\rm even},   \; a' : {\rm odd}  \rightarrow    \sum_{b< 2^{m'}} \psi_{b,a} \, \psi_{b, a'} = \delta_{a,2} \,  \psi_{0, a'}.
\label{17}
\eeq
Eq. (\ref{3}) implies that $\psi_{0, a}=1$  if $a$ is a prime and $\psi_{0, a}=0$ if  it is not. 
So $\psi_{0,a}$ is equal to the characteristic function of prime numbers  $\chi_\pi(a)$. 
We have then derived

 \beq
m> 1, \; a : {\rm even},   \; a' : {\rm odd}      \rightarrow \rho^A_{a,a'}  =  \frac{ \delta_{a,2}  \, \chi_{\pi}(a')}{ \pi(N)}. 
\label{20}
\eeq
Finally, in order to compute $\rho^A_{a,a'}$,  when $a \neq a'$ are both odd,
we shall use the  counting function $\pi_{a;  b, b'}(x)$ defined by Eq.(\ref{pab}). 
The sum in the expression Eq. (\ref{6})
for $\rho^A_{a,a'}$, when $a <  a'$ are both odd,  
is the number of pairs of odd primes of the form 
$(p, p')  = (a + 2^m b, a' + 2^m b)$,  with  $p \leq N$ and $p' \leq N$. 
This is just the counting function  $\pi_{a;b, b'}(x)$ with 
$(x, a, b, b')  \rightarrow (N, M, a, a')$, hence 

 \beq
m> 1, \; a : {\rm odd},   \; a' : {\rm odd}      \rightarrow \rho^A_{a,a'}  =  \frac{ \pi_{M;  a, a'}( N) }{ \pi(N)}. 
\label{21}
\eeq
Collecting  the previous results in a block form (even, odd) one finds 

\beq
\rho_A =  \frac{1}{ \pi(N)} 
\left( 
\begin{array}{cc}
\delta_{a,2} \delta_{a', 2} & \delta_{a,2}  \, \chi_{\pi}(a') \\ 
\delta_{a',2} \,  \chi_{\pi}(a) & \delta_{a, a'} \pi_{M,a}(N) + ( 1 - \delta_{a, a'}) \pi_{M;  a, a'}( N) \\
\end{array}
\right). 
\label{22}
\eeq

As an example, let us choose a system with $n=6$ qubits. The prime state
in Eq. (\ref{1}) contains  $\pi(2^6) = 18$ terms corresponding to all the 
primes  less that $N = 64$. We split the system into  two blocks 
with $m = m'=3$ qubits each. The binary representation of the primes
yields the function $\psi_{b,a}$ of Eq.(\ref{3}) which we write
in matrix form as

\beq
\psi = \left( 
\begin{array}{cccc|cccc}
0 &  1 &  0 &  0 &  0 &  1 &  1 &  1  \\
0 &  0 &  0 &  0 &  1 &  1 &  0 &  1 \\
0 &  0 &  0 &  0 &  0 &  0 &  1 &  0 \\  
0 &  0 &  0 &  0 &  0 &  0 &  1&  0 \\
\hline 
0 &  0 &  0 &  0 &  0 &  1 &  1 &  0 \\
0 & 0 & 0 &  0 &  0 &  0 &  1 &  1 \\
0 & 0 &  0 &  0 &  1 &  1 &  0 &  1 \\
0 &  0 &  0 &  0 &  0 &  1 &  1 &  0 \\
 \end{array}
\right) \leftrightarrow 
\left( 
\begin{array}{cccc|cccc}
0 &  2 &  0 &  0 &  0 &  3 &  5 &  7  \\
0 &  0 &  0 &  0 &  17 &  19 &  0 &  23 \\
0 &  0 &  0 &  0 &  0 &  0 &  37 &  0 \\  
0 &  0 &  0 &  0 &  0 &  0 &  53 &  0 \\
\hline 
0 &  0 &  0 &  0 &  0 &  11 &  13 &  0 \\
0 & 0 & 0 &  0 &  0 &  0 &  29 &  31 \\
0 & 0 &  0 &  0 &  41 &  43 &  0 &  47 \\
0 &  0 &  0 &  0 &  0 &  59 &  61 &  0 \\
 \end{array}
\right)
\label{23}
\eeq
where the rows and columns corresponds to the indices $b$ and $a$ respectively
which are ordered in the even-odd sequence: $(0,2,4,6,1,3,5,7)$. The matrix on the RHS
gives the associated  prime numbers, for example $\psi_{4,5}=1$
corresponds   to $p = 5 + 2^3 \times 4 = 37$.  The density matrix $\rho_A$
can be readily computed as the product

\beq
\rho_A = \frac{1}{\pi(2^6)}  \psi^\dagger \, \psi  = \frac{1}{18} \left( 
\begin{array}{cccc|cccc}
0 &  0 &  0 &  0 &  0 &  0 &  0 &  0  \\
0 &  1 &  0 &  0 &  0 &  1 &  1 &  1 \\
0 &  0 &  0 &  0 &  0 &  0 &  0 &  0 \\  
0 &  0 &  0 &  0 &  0 &  0 &  0 &  0 \\
\hline 
0 & 0 &  0 &  0 &  2 &  2 &  0 &  2 \\
0 & 1 & 0 &  0 &  2 &  5 &  3 &  3 \\
0 & 1 &  0 &  0 &  0 &  3 &  6 &  2 \\
0 & 1 &  0 &  0 &  2 &  3 &  2 &  4 \\
 \end{array}
\right). 
\label{24}
\eeq
Comparing this matrix with the expression  in Eq. (\ref{22})
one  verifies the  even-even, even-odd and odd-even  blocks, the latter ones
following from the identities  $\chi_{\rm prime}(i) =1$ for $i=3,5,7$. The diagonal entries
of the odd-odd block are given by  the number of primes in the arithmetic progressions  $8 k + a$
with $a=1,3,5,7$, 

\ba
\pi_{8, 1}(64) = 2  & : &  \left\{ 17, 41 \right\} 
\label{25} \\
\pi_{8, 3}(64) = 5  & : &  \left\{ 3, 11, 19, 43, 59  \right\}  \nonumber \\
\pi_{8, 5}(64) = 6  & : &  \left\{ 5, 13, 29, 37,  53, 61  \right\}  \nonumber \\
\pi_{8, 7}(64) = 4  & : &  \left\{ 7, 23, 31,  47  \right\}  \nonumber 
\ea
Similarly, the off-diagonal entries of the odd-odd block counts the number 
of pairs in the associated arithmetic progressions:

\ba
\pi_{8; 1,3}(64) = 2 & : & \left\{ (17,19), (41, 43) \right\} \label{26} \\
\pi_{8; 1,5}(64) = 0 & : & \left\{  \right\} \nonumber  \\
\pi_{8; 1,7}(64) = 2 & : & \left\{ (17,23), (41, 47) \right\} \nonumber \\
\pi_{8; 3,5}(64) = 3 & : & \left\{ (3,5), (11, 13), (59, 61) \right\} \nonumber \\
\pi_{8; 3,7}(64) = 3 & : & \left\{ (3,7), (19, 23), (43, 47) \right\} \nonumber \\
\pi_{8; 5,7}(64) = 2 & : & \left\{ (5,7), (29, 31) \right\} \nonumber 
\ea
Notice that $17$ and $13$ belong to the  series $1 \, {\rm mod} \,  8$ and $5 \,  {\rm mod} \,  8$ respectively,  
and that  $17 = 8 \times 2 +1$ and $13 = 8 \times 1 + 5$. Hence they do not  satisfy the definition in Eq. (\ref{pab}),
and therefore do not contribute to  $\pi_{8; 1,5}(64)$ which is actually zero. 

\section*{Appendix A2: Asymptotic limit of prime counting functions} 

In this appendix we present numerical results concerning the asymptotic
behavior of several prime number  functions that appear in the main text.
Some of the results are well known but others are new,  as far
as we know.

\subsection{The prime counting function $\pi(N)$} 

\begin{figure}
\centering
\includegraphics[width=0.45\textwidth]{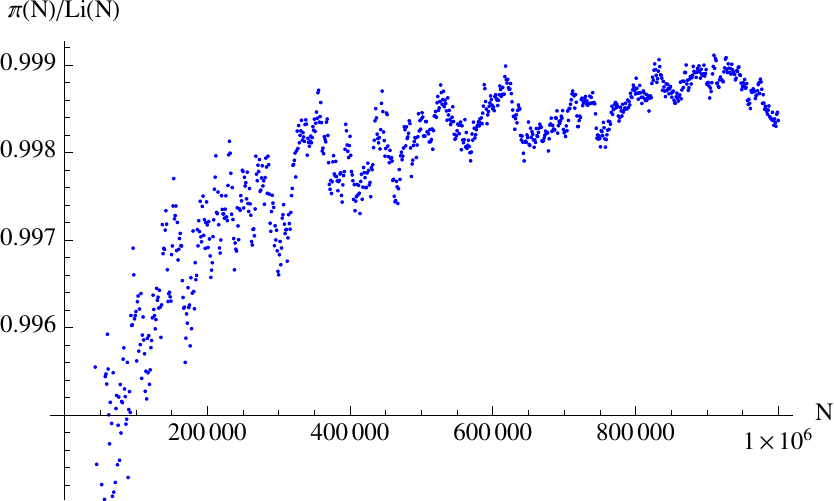}
\hspace{0.1cm}
\includegraphics[width=0.45\textwidth]{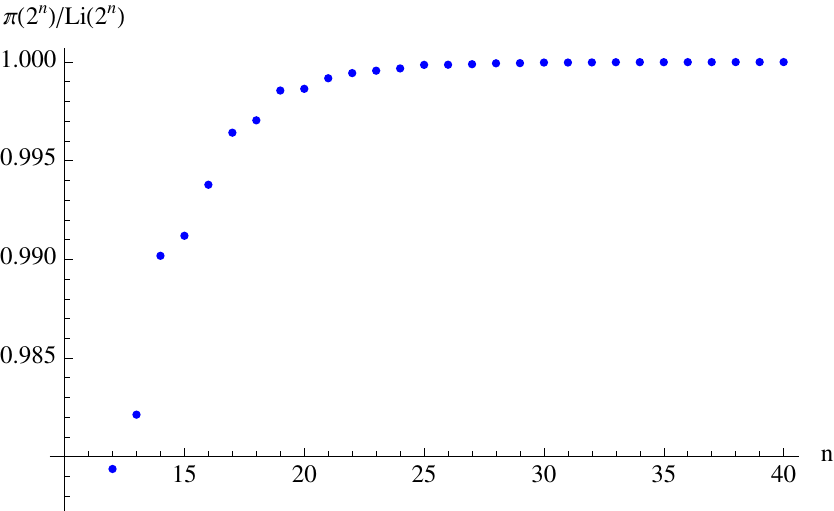}
\caption{Numerical verification  of the PNT in Eq. (\ref{ap1}). 
Left:  $N =10^3 - 10^6$ in steps of $10^3$. Right:  $N = 2^n$ for $n=10- 40$. 
 } 
\label{pnt} 
\end{figure}

The PNT given in Eq. (\ref{PNT}) is stated more precisely as the  limit

\beq
\lim_{N \rightarrow \infty} \frac{ \pi(N)}{ Li(N)} = 1. 
\label{ap1}
\eeq
Fig. \ref{pnt} shows the convergence of the ratio $\pi(N)/Li(N)$ for two examples. 
In the case where $N= 2^n$  the fluctuations are not  as visible as in the first case, but they  lie
within the limits imposed  by the Riemann hypothesis,  that are of  order $N^{-1/2}  \log^2 N$ as shown in Eq. (\ref{Riemannbound}). 
From Fig.(\ref{pnt}) one would be tempted to conjecture that
  $\pi(N) < Li(N)$, for all $N$.  However, Littlewood showed in 1914  that the difference $\pi(N)- Li(N)$ 
 changes infinitely often.  It is not known the  lowest value of $N$, for which $\pi(N) > Li(N)$.
 Skewes gave a first  estimate in terms of the  {\em googolian}  number $10^{10^{10^{34}}}$, known as the Skewes number, 
 but this upper bound  value has been reduced to a {\em modest}   $ e^{729} \sim 3.98 \times 10^{316}$.

\subsection{The prime counting function for arithmetic series $\pi_{a,b}(N)$}

\begin{figure}[h]
\centering
\includegraphics[width=0.45\textwidth]{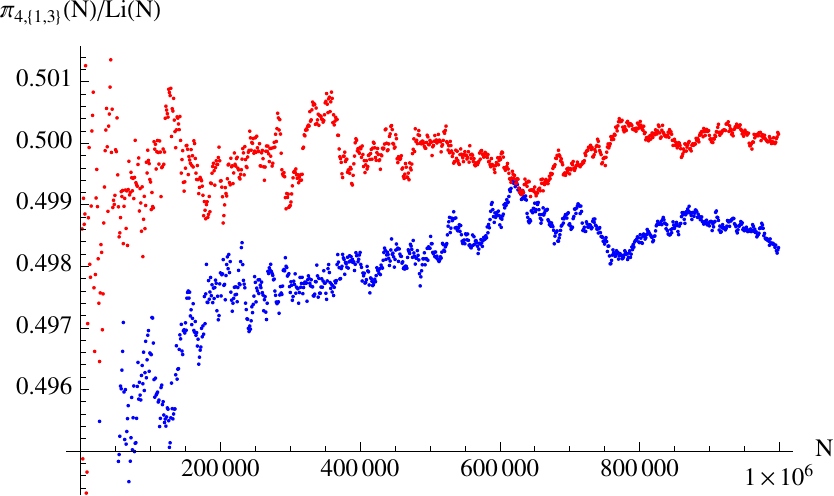}
\hspace{0.1cm}
\includegraphics[width=0.45\textwidth]{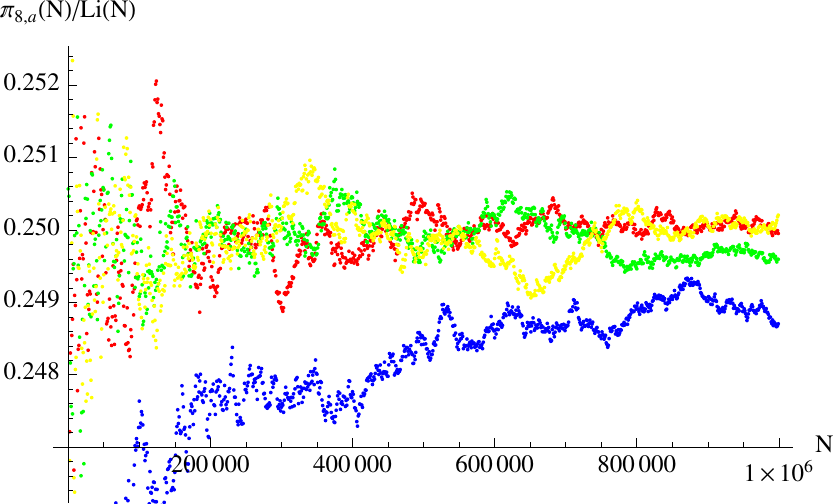}
\caption{Left: $\pi_{4, b}(N)$ for $b=1$ (blue color on line) and $b=3$ (red color  on line). 
Right: $\pi_{8, b}(N)$ for $b=1$ (blue), $b=3$ (red), $b=5$ (green) and
$b=7$ (yellow). The curves converge towards $1/\phi(a) = 1/2$ and $1/4$,
respectively, in agreement with Eq.(\ref{ap2})
 } 
\label{pnt2} 
\end{figure}

The PNT for arithmetic progressions,  Eq.(\ref{32}),  amounts to

\beq
\lim_{N \rightarrow \infty} \frac{ \pi_{a,b}(N)}{ Li(N)} = \frac{ 1}{\phi(a)},  \qquad gcd(a,b) = 1, 
\label{ap2}
\eeq
and is illustrated in Fig. \ref{pnt2} in the cases $\pi_{4, b}(N)$ and $\pi_{8,b}(N)$. 
Notice that $\pi_{4,3}(N)- \pi_{4,1}(N)$ is mostly positive. This  difference is 
known as the Chebyshev bias and changes infinitely often. Fig. \ref{pnt2}  shows
one crossing. Similar pattern  occurs  for $\pi_{8, b}(N)$.

\subsection{The prime correlation  function $\pi_2(k,N)$}

\begin{figure}[h]
\centering
\includegraphics[width=0.4\textwidth]{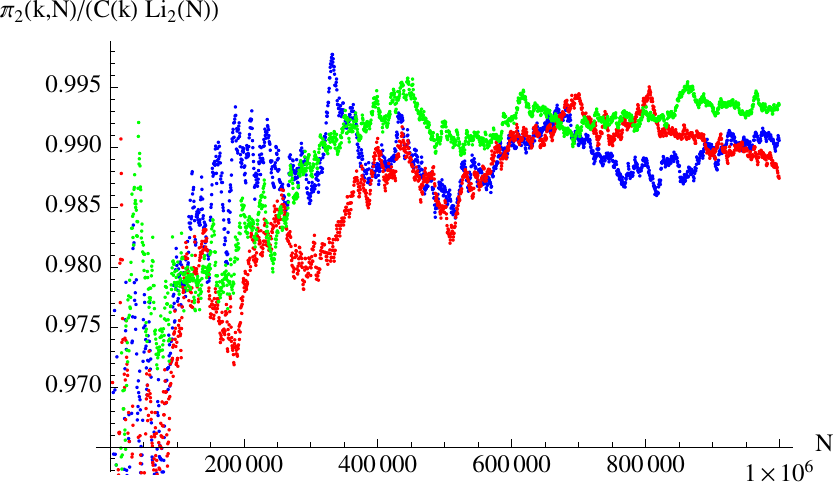}
\caption{Illustration of Eq.(\ref{ap3})   for $k=2,4,6$.  
 } 
 \label{pnt3} 
 \end{figure}

The Hardy-Littlewood conjecture of Eq. (\ref{34}) reads

\beq
\lim_{N \rightarrow \infty} \frac{ \pi_{2}(k, N)}{ Li_2(N)} = C(k), 
\label{ap3}
\eeq
where $C(k)$ is defined in Eq.(\ref{35}). Fig. \ref{pnt3} depicts this limit for $k=2,4,6$, where
$C(k)$ takes the following values

\beq
C(2) = 1.32032.., \quad C(4) = C(2), \quad C(6) = 2 C(2). 
\label{ap4}
\eeq

\vspace{0.5 cm} 

\subsection{The arithmetic correlation  function $\pi_{M, a, a'}(N)$}

\begin{figure}[h]
\centering
\includegraphics[width=0.3\textwidth]{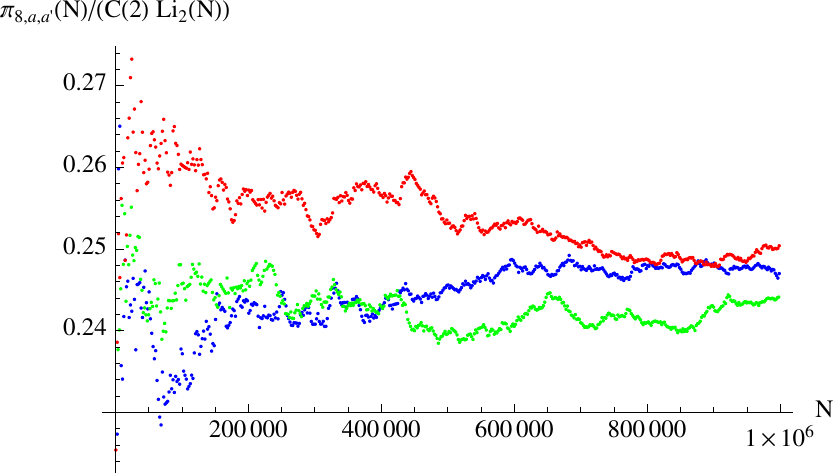}
\hspace{0.1cm}
\includegraphics[width=0.3\textwidth]{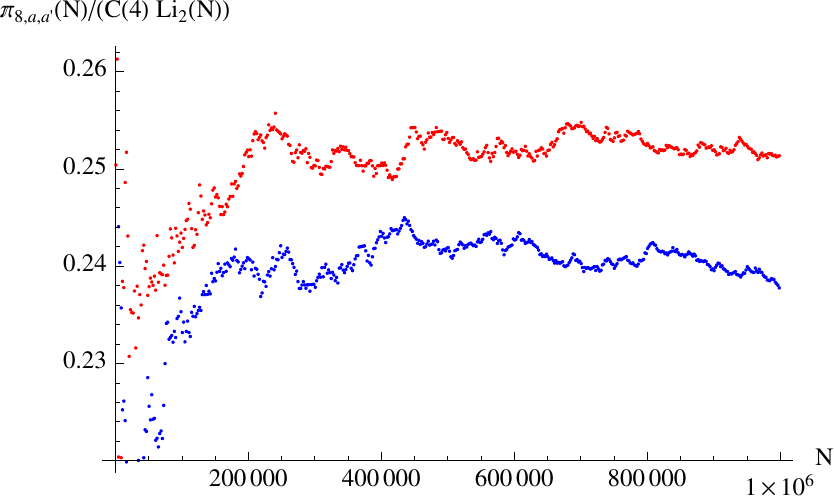}
\hspace{0.1cm}
\includegraphics[width=0.3\textwidth]{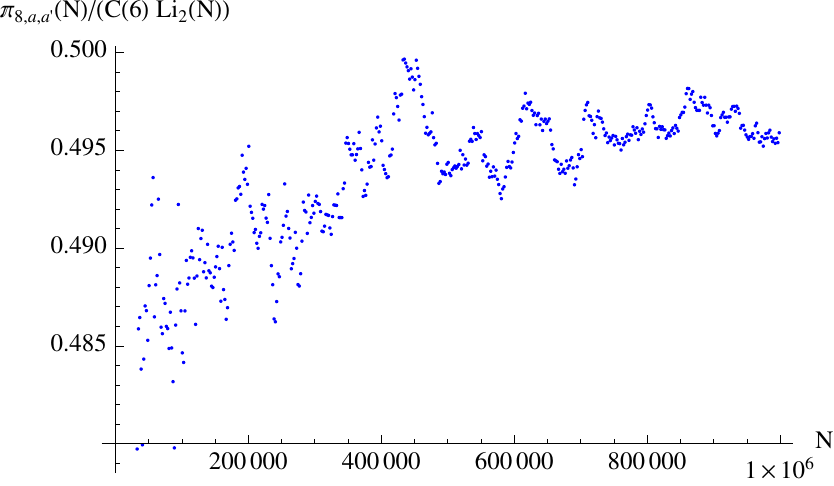}
\caption{Illustration of Eq.(\ref{ap5})   for $M=8$. Left: $(a, a') = (1,3), (3,5), (5,7)$ (color on line: blue, red, green). 
Center: $(a,a') = (1,5), (3,7)$ (color on line:  blue, red).
Right: (a, a') = (1,7) (color on line: blue).  Note that $C(2) = C(4)= C(6)/2$ and $\phi(8)=4$, which explains the asymptotic values
of the abcissas. 
 } 
\label{pnt4} 
\end{figure}

Eq. (\ref{36}) means
\beq
\lim_{N \rightarrow \infty} \frac{ \pi_{M; a, a'}(N)}{Li_2(N)} = \frac{ C(|a- a'|)}{\phi(M)} 
\label{ap5}
\eeq
and it is illustrated in Fig. \ref{pnt4}  for $M=8$.

\section*{Appendix A3: The Hardy-Littlewood constants  $C(k)$} 

\begin{figure}[h]
\centering
\includegraphics[width=0.35\textwidth]{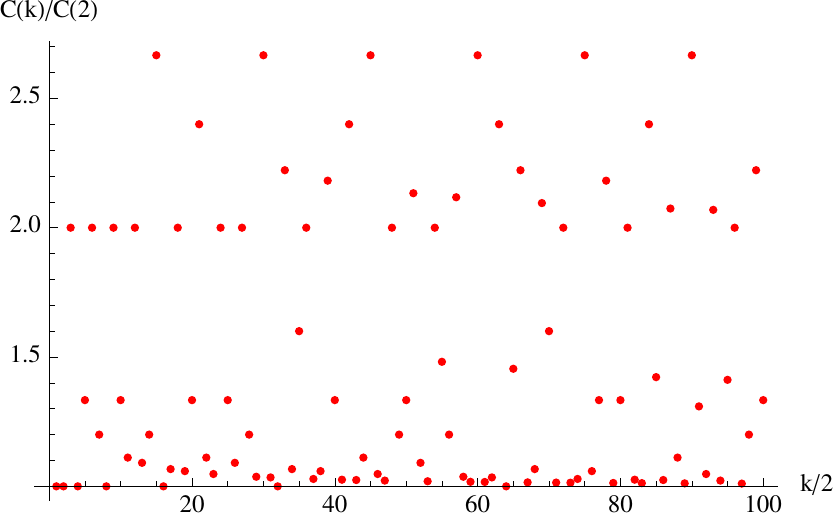} 
\hspace{0.3cm}
\includegraphics[width=0.35\textwidth]{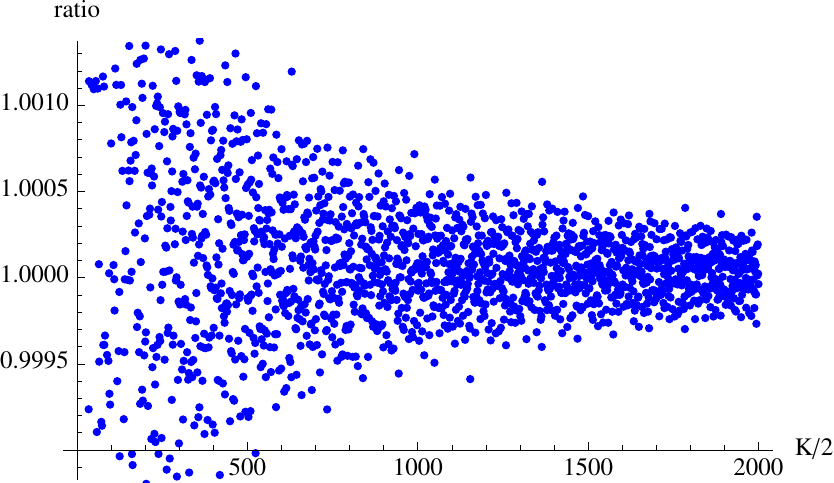}
\caption{Left: $C(k)/C(2)$ for $k = 2, 4, \dots, 200$.  Right: ratio = $\sum_{k=1}^K C(k) /(K - \frac{1}{2} \log K)$
for $K$ up to 4000. 
 } 
\label{Ck200}  
\end{figure}

The constants $C(k)$ that appear in the Hardy-Littlewood conjecture in Eq. (\ref{35}) 
vary irregularly with $k$. The behavior  is shown in   Fig.\ref{Ck200}-left. 
However, the sum of $C(k)$ satisfies an asymptotic  formula obtained by Keating  \cite{K93,BK99}(see  in Fig.\ref{Ck200}-right),

\beq
\sum_{k=1}^K C(k) \sim K - \frac{1}{2} \log K , \qquad K >> 1, 
\label{k1}
\eeq
which  implies the average asymptotic behavior

\beq
\langle C(k) \rangle  \sim 1 -  \frac{1}{2 |k|}, \qquad k  \gg  1.
\label{k2}
\eeq

The approximation in Eq. (\ref{k2}) played a  crucial role in the derivation of the statistics of the Riemann
zeros that is described by the Gaussian Unitary Ensemble (GUE), as conjectured by Montgomery 
in the 70's and confirmed numerically by Odlyzko in the 80's. In our study of the entanglement of the Prime state, instead
of Eq.(\ref{k1}) we need the asymptotic behavior of ${\rm Tr} \, {\bf C}_m^\alpha$ for $\alpha =2, 3, \dots$. 
In  the case  $\alpha =2$, that is   ${\rm Tr} \, \bar{\rho}_A^2$, we need the asymptotic 
behavior of  $\sum_{a,a'}^K  C(|a - a'|)^2$, that will be computed below using  the probabilistic
methods that lead to Eq. (\ref{k1}). To do so,  we  shall first review 
the probabilistic properties of the primes and the division of integers by primes (we follow closely reference \cite{K93}).

\subsection{Asymptotic behaviour  of $\sum_{k=1}^K C(k)$} 

The PNT (\ref{PNT}) implies that the probability $P(n)$ that $n$ is a prime behaves as 

\beq
P(n) \sim \frac{1}{ \log n}, \qquad n \rightarrow \infty.
\label{k3}
\eeq
The Hardy-Littlewood conjecture of Eq. (\ref{34}) means that the probability that both $n$ and $n-m$ are primes is,
in the limit $n \rightarrow \infty$,

\beq
P(n, n-m) \sim P^2(n) \, \alpha(m), \qquad n \rightarrow \infty.
\label{k4}
\eeq
where $\alpha(m)$ coincides with $C(m)$ (in this appendix we use the notation of reference \cite{K93}).  
The probabilistic concepts can  be also applied to divisibility of integers. For example, of the positive
integers a fraction $1/p$  are divisible by a prime. Hence,  for a large integer $n$ the probability that
a prime $p$ divides $n$  is given by $1/p$. Similarly,  the 
probability that an integer $n$ is divided by $p$ (i.e. $p|n$),  but not by $p^2$ (i.e. $p^2 {\slashed |} n$)  is $1/p - 1/p^2 = (p- 1)/p^2$. 
The fundamental theorem of Arithmetics imply that division by different primes are independent
operations , so from the probabilistic viewpoint the primes are statistically independent variables.
Applying these probabilistic techniques one can derive $\alpha(m)$, 
\beq
\alpha(m) = \alpha \prod_{p >  2, \,  p | m}  
\left( 1 + \frac{1}{p-2} \right) \;  (m : {\rm even}), \qquad \alpha(m) = 0 \; (m : {\rm odd}), 
\label{k5}
\eeq
where $\alpha/2$ is the twin prime constant
\beq
\alpha  =  2 \prod_{p >  2}  
\frac{p (p-2)}{(p-1)^2} = 1.3203...
\label{k6}
\eeq

Let us now factorize $m$ into primes

\beq
m = 2^{b_0} p_1^{a_1} \dots p_s^{a_s}  \quad (p_i > 2,  a_i \geq 1, b_0 \geq 1  ),  
\label{k7}
\eeq
that  yields 

\barray 
\alpha(m) &  =  & \alpha \left( 1 + \frac{1}{p_1-2} \right) \dots  \left( 1 + \frac{1}{p_s-2} \right)
\label{k8} \\
& = & \alpha \left(  1 + \sum_{1 \leq i \leq s}   \frac{1}{p_i-2}  + \sum_{1 \leq i_1 < i_2 \leq s}   \frac{1}{(p_{i_1}-2 ) (p_{i_2}-2 )  }  
+ \dots +   \frac{1}{(p_{i_1}-2 )  \dots (p_{i_s}-2 )}
\right). 
\nonumber 
\earray 
A generic term in the sum can be written as

\beq
\beta(d) = \prod_{p >2, p |d} \frac{1}{ p -2} = 
 \frac{1}{(p_{i_1}-2 )  \dots (p_{i_l}-2 )}, \quad  {\rm with} \; d = p_{i_1} \dots p_{i_l},
\label{k9}
\eeq
so that Eq.(\ref{k8}) involves all the $d's$ containing some of the  $p's$ of the integer $m$, but only once. 
Hence $d$ are divisors of $m$, with the latter properties. In the rest of the cases one writes 

\beq
\beta(1) = 1, \qquad 
\beta(d) = 0 \quad {\rm if} \; d: {\rm even} \; \;  {\rm or } \; \; p_i^2 | d.
\label{k10}
\eeq 
After these manipulation  Eq. (\ref{k5}) becomes

\beq
\alpha(m) = \alpha \sum_{d | m}   \beta(d) \; \;  (m: {\rm even}) 
\label{k11}
\eeq
The next goal  is to compute  the following sum in the limit $X \rightarrow \infty$

\beq
\sum_{m=1}^X \alpha(m) = \sum_{i=1}^{[X/2]}  \alpha(2i) =\alpha \sum_{i=1}^{[X/2]}    \sum_{d | 2 i }   \beta(d) =\alpha \sum_{i=1}^{[X/2]}    \sum_{d |  i }   \beta(d).  
\label{k12}
\eeq
In this expression  $[ x ]$ denote the integer part of $x$.  The last equality comes from the fact that 
if $d$ is even, then  $\beta(d)=0$, hence the condition $d |2i$ amounts to $d | i$ and this happens
$[X/( 2 d)]$ times so,

\beq
\sum_{m=1}^X \alpha(m) = \alpha \sum_{d \leq X/2}  \beta(d) \left[ \frac{ X}{2 d} \right].
\label{k13}
\eeq
One can verify Eqs.(\ref{k11}) and (\ref{k13}) using the data provided in  Table IV.

\begin{center}
\begin{tabular}{|c|ccccccc|}
\hline
$m $  & 2 &   4  & 6 & 8 & 10  & 12 & 14 \\
$\alpha(m)$  & 1 & 1 & 2 & 1 & $\frac{4}{3}$ & 2 & $\frac{6}{5}$   \\
\hline 
$X$ & 2  & 4  & 6 & 8 & 10 & 12 & 14   \\
$\sum_{m=1}^X \alpha(m)$ & 1 & 2 & 4 & 5 & $\frac{19}{3}$  & $\frac{25}{3}$ & $\frac{143}{15}$ \\
\hline 
$d$ & 1  & 3  & 5 & 7 & 11 & 13 & 15   \\
$\beta(d)$ & 1  & 1  & $\frac{1}{3}$ & $\frac{1}{5}$ & $\frac{1}{9}$ & $\frac{1}{11}$ & $\frac{1}{3}$   \\
\hline
\end{tabular}

\vspace{0.2 cm}

TABLE IV:  Some numerical values of the functions $\alpha(m)$, $\sum_{m=1}^X \alpha(m)$ and $\beta(d)$. 
\end{center}

Any real number $x$ can be decomposed as $x = [x] + \{x \}$, where $\{x \}$ denote
the fractional part. So Eq. (\ref{k13}) splits as

\beq
\sum_{m=1}^X \alpha(m) =  \frac{\alpha X}{2}  \sum_{d \leq X/2}  \frac{ \beta(d)}{d}
-  \alpha \sum_{d \leq X/2}  \beta(d) \left\{  \frac{ X}{2 d} \right \}.
\label{k14}
\eeq
The problem now is to estimate the RHS of this equation  using probabilistic arguments. 
Eq.(\ref{k9}) shows that $\beta(d)$ behaves on average as $1/d$, 

\beq
\beta(d) \sim 
 \frac{1}{ p_{i_1}   \dots  p_{i_l} } = \frac{1}{d}, \quad  {\rm with} \; d = p_{i_1} \dots p_{i_l}. 
\label{k15}
\eeq
Hence one can  consider  the expectation value of the product $\langle d \beta(d) \rangle$
for large $d$. If $d$ is even, one gets 0, so there is an overall  factor $1/2$. If $d$ is odd
there are two situations concerning the contribution of a prime $p$ to the product $d \beta(d)$.
If $p$ does not divide $d$, it contributes with a factor 1 to $d \beta(d)$, with probability $(p-1)/p$. 
If $p$ divides $d$, but $p^2$ does not, the contribution to the product is $p \times 1/(p-2)= p/(p-2)$
with a probability $(p-1)/p^2$. These arguments leads to 

\beq
\langle d \beta(d) \rangle \simeq \frac{1}{2} \prod_{p > 2} \left[  \left( 1 - \frac{1}{p} \right) 
+ \frac{p}{p-2} \left(  \frac{p-1}{p^2} \right)  \right]  = \frac{1}{\alpha}, 
\label{k16}
\eeq
where one has included all the primes that gives a convergent product. 
The conclusion is that $\beta(d)$ behaves for large $d$  as $1/(\alpha d)$, 
and therefore $\sum_{d \leq X/2} \beta(d)/d \sim \sum_{d \leq X/2} 1/( \alpha d^2)$ is convergent
in the limit $X \rightarrow \infty$ towards the value 

\beq
\sum_{d=1}^\infty \frac{\beta(d)}{d}  
= 1+ \sum_{d=3}^\infty  \prod_{p>2, p |d} \frac{1}{p(p -2)} =  \prod_{p>2} \left( 1+  \frac{1}{p(p -2)} \right)  = \frac{2}{\alpha}.
\label{k17}
\eeq 
This constant cancels the factor $\alpha/2$ of the first term in Eq. (\ref{k14}), that becomes simply $X$. 
The second sum in Eq. (\ref{k14}) can be approximated assuming that $\left\{  \frac{ X}{2 d} \right \}$
varies uniformly in the range $[0, 1]$, so its average is $1/2$ yielding in the limit  $X \rightarrow \infty$

\beq
 \sum_{d \leq X/2}  \beta(d) \left\{  \frac{ X}{2 d} \right \} 
\sim  \frac{1}{2}  \sum_{d \leq X/2}  \beta(d) \sim  \frac{1}{2 \alpha }  \sum_{d \leq X/2}  \frac{1}{d} \sim \frac{1}{2 \alpha} 
\left(  \log \frac{X}{2} +   \gamma \right), 
\label{k18}
\eeq
where $\gamma$ is the Euler's constant.  The final result is (see Eq. (\ref{k1}))

\beq
\sum_{m=1}^X \alpha(m)  \sim X - \frac{1}{2}  \log X, 
\label{k19}
\eeq
where we dropped a constant term.  From here one can conclude that 
the average behavior of $\alpha(m)$ is then given by (recall Eq. (\ref{k2})) 

\beq
\langle \alpha(x) \rangle  \sim     1  - \frac{1}{2 x}, \qquad x \rightarrow \infty.
\label{k191}
\eeq

\subsection{Asymptotic behaviour  of $\sum_{a, a'}^K C^2(|a -a'|)$} 

The asymptotic behavior of the purity in Eq. (\ref{37})  requires the 
evaluation of the following sum for large values of $d= 2^{m-1}$

\beq
\sum_{i, j=1}^d C^2( 2|i-j|). 
\label{k200}
\eeq
Defining $d = X/2$,   we can write this sum as

\beq
\sum_{i, j=1}^{X/2} \alpha^2( 2 |i-j|)  =   \sum_{i=1}^{X/2}  ( X  - 2  i )  \alpha^2( 2 i)   = X   \sum_{m=1}^X \alpha^2(m) -
\sum_{m=1}^X m \, \alpha^2(m). 
\label{k201}
\eeq
We shall consider separately the sums  $\sum_{m=1}^{X}  \alpha^2(m)$ and $\sum_{m=1}^{X} m \,   \alpha^2(m)$. 
Using Eq. (\ref{k11}) we write

\barray 
\sum_{m=1}^{X}  \alpha^2(m)  & = &  \alpha^2 \sum_{i=1}^{[X/2]}  \sum_{d_1 | i} \sum_{d_2 |i} \beta(d_1) \beta(d_2).
\label{k22}
\earray
$d_1,  d_2$  divide $i$, hence the least common multiple of these numbers, ${\rm lcm}(d_1, d_2)$,
must divide $i$,  so 

\beq
\sum_{i=1}^{[X/2]} 1 |_{{\rm lcm}(d_1, d_2) | i}   = \left[ \frac{ X}{ 2 \,  {\rm lcm}(d_1, d_2)} \right], 
\label{k221}
\eeq
and then 

\barray 
\sum_{m=1}^{X}  \alpha^2(m)  & = & 
\alpha^2 \sum_{d_1, d_2  \leq X/2}  \beta(d_1) \beta(d_2) \left[ \frac{ X}{ 2 \,  {\rm lcm}(d_1, d_2)} \right]  
\label{k31}  \\ 
& = & 
\frac{\alpha^2 X}{2}  \sum_{d_1,  d_2  \leq X/2}   \frac{ \beta(d_1) \beta(d_2)  }{   {\rm lcm}(d_1, d_2)}  -
\alpha^2 \sum_{d_1,  d_2  \leq X/2}  \beta(d_1) \beta(d_2) \left\{  \frac{ X}{ 2 \,  {\rm lcm}(d_1, d_2)} \right\} . 
\nonumber 
\earray
The first sum is convergent in the limit $X \rightarrow \infty$  and is given by

\beq 
\sum_{d_1,  d_2 =1}^\infty    \frac{ \beta(d_1) \beta(d_2)  }{   {\rm lcm}(d_1, d_2)}  =
\prod_{p>2} \left( 1 + \frac{2}{ p ( p-2) } + \frac{1}{p (p-2)^2} \right) \equiv \frac{2}{\alpha_2}.  
\label{k311}
\eeq
This formula includes  all the  contributions where $d_1$ and $d_2$ are either 1, or 
odd square-free integers. 
For example,  if $d_1$ and $d_2$ have the factorizations

\beq
d_1 = \prod_i \,  p_i \, \prod_j  \, q_j, \quad d_2 = \prod_ i \,  p_i \, \prod_k \, r_k, \quad {\rm lcm}(d_1, d_2) = \prod_i \, p_i \, 
\prod_ j \,  q_j \, \prod_k \, r_k
 \qquad p_i  \neq q_j \neq r_k (\neq p_i)
\label{k32}
\eeq
one finds 

\beq
 \frac{ \beta(d_1) \beta(d_2)  }{   {\rm lcm}(d_1, d_2)}  =  \prod_i \frac{1}{p_i  (p_i -2)^2}   \prod_j \frac{1}{q_j  (q_j -2)} 
 \prod_k \frac{1}{r_k  (r_k -2)}. 
 \label{k33}
 \eeq
 In Eq.(\ref{k311}), the term $1/( p (p-2)^2)$ arises from the primes $p$ common to  $d_1$ and $d_2$, while
 the term $1/(p(p-2))$, arises from those that are not common. 
 The constant $\alpha_2$ ,  defined in Eq.
(\ref{k311}),  is related to the twin prime constant $\alpha$   as

\beq
\frac{\alpha^2}{\alpha_2} =  2 \prod_{p>2} \left( 1 + \frac{1}{(p-1)^3} \right)
= 2.30096... 
\label{k331}
\eeq
Hence the first term in Eq. (\ref{k31})
behaves in the limit $X \rightarrow \infty$ as $\alpha^2/\alpha_2 X$. 
Let us now consider the second term in Eq. (\ref{k31}). Here we make the  approximation 
that led to Eq. (\ref{k18}), which is to  assume  that $ \left\{  \frac{ X}{ 2 \,  {\rm lcm}(d_1, d_2)} \right\}$
varies uniformly in the range $[0, 1]$, so its average is $1/2$ 
\begin{figure}[h]
\centering
\includegraphics[width=0.3\textwidth]{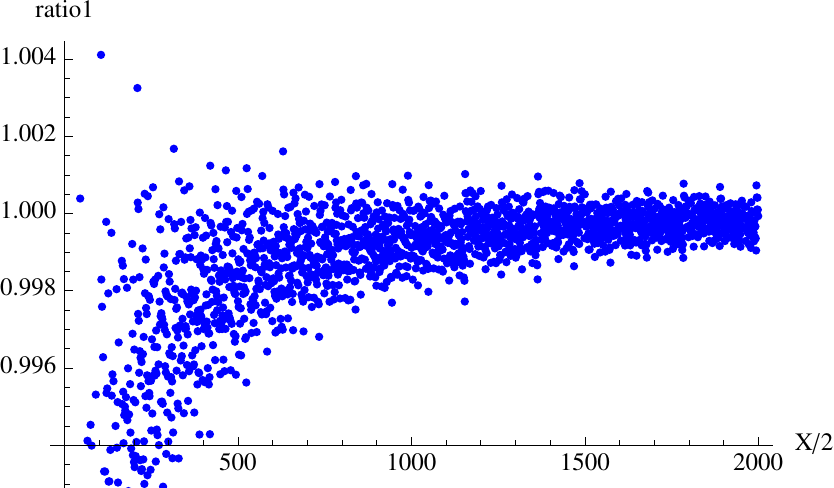} 
\hspace{0.5cm}
\includegraphics[width=0.3\textwidth]{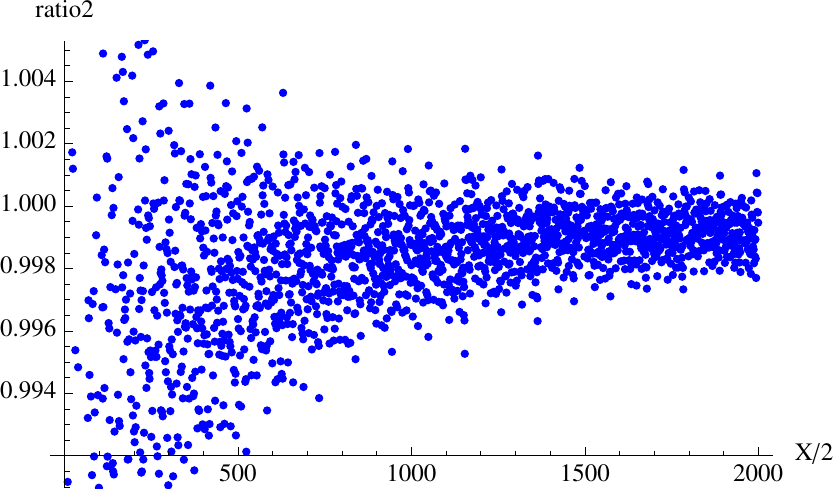}
\hspace{0.5cm}
\includegraphics[width=0.3\textwidth]{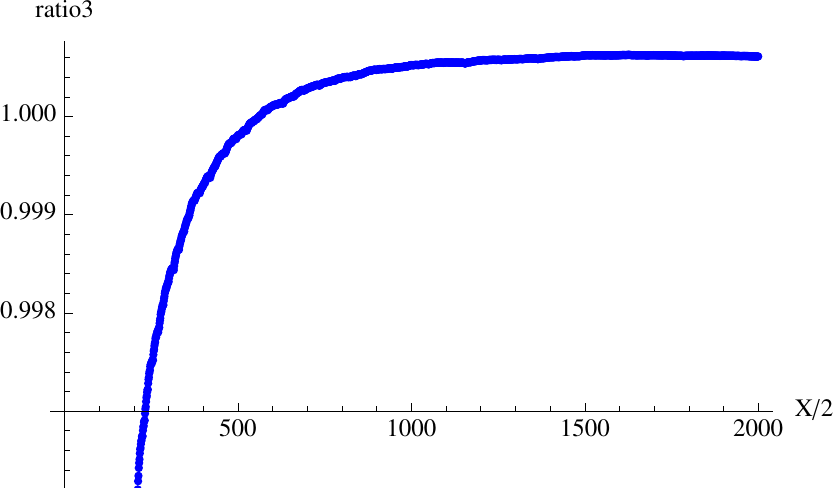}
\caption{Left:  $ratio_1 = \sum_{m=1}^X \alpha^2(m)/ ( \frac{\alpha^2}{\alpha_2} X - \frac{1}{2} (\log X)^2)$
(see Eq. (\ref{k333})). 
Center: $ratio_2 = \sum_{m=1}^X m  \, \alpha(m)^2/ ( \frac{\alpha^2}{2\alpha_2} X^2)$
(see Eq. (\ref{k361})). 
Right:  $ratio_3  = \sum_{i, j=1}^{X/2} \alpha^2( 2 |i-j|) / ( \frac{\alpha^2}{2 \alpha_2} X - \frac{X}{2} (\log X)^2)$
(see Eq. (\ref{k37}))
In all  cases   $X/2$ runs up to 2000. 
 } 
\label{sumC22}  
\end{figure}

\beq
 \sum_{d_1,  d_2  \leq X/2}  \beta(d_1) \beta(d_2) \left\{  \frac{ X}{ 2 \,  {\rm lcm}(d_1, d_2)} \right\} \sim
 \frac{1}{2}  \sum_{d_1,  d_2  \leq X/2}  \beta(d_1) \beta(d_2) \sim \frac{1}{2 \alpha^2} \left(  \log   X \right)^2, 
 \label{k332}
\eeq
where we have used Eq. (\ref{k16}). Collecting the previous results we get (see Fig. \ref{sumC22}-left for a numerical confirmation)

\beq
\sum_{m=1}^X \alpha^2(m) \sim   \frac{ \alpha^2}{\alpha_2} X - \frac{1}{2} \left(  \log   X \right)^2. 
\label{k333}
\eeq
This result is quite different from the one obtained replacing $\alpha^2(m)$ by  $\langle \alpha(m) \rangle^2$
 using Eq. (\ref{k191}), 

\beq
\sum_{m=1}^X \langle \alpha(m) \rangle^2   \sim   X - \log X +  {\rm cte},
\label{k3331}
\eeq
which is a consequence of the fluctuations of $\alpha(m)$. In other words, the sum in Eq. (\ref{k333}),
 provides further information about the distribution of prime numbers. 
 
 Let us now consider the second term of  Eq. (\ref{k201}), that using Eq. (\ref{k11}) becomes

\barray 
\sum_{m=1}^{X}  m \,  \alpha^2(m)  & = & 2  \alpha^2 \sum_{i=1}^{[X/2]}  \sum_{d_1 | i} \sum_{d_2 |i}  i \,  \beta(d_1) \beta(d_2).
\label{k35}
\earray

Similarly  to Eq. (\ref{k221}), one finds 
\barray 
\sum_{i=1}^{[X/2]} i  |_{{\rm lcm}(d_1, d_2) | i}   &  = &   \frac{\ell_{12}}{2}  \left( 1+ \left[ \frac{ X}{ 2 \, \ell_{12} } \right]  \right) 
 \left[ \frac{ X}{ 2 \,  \ell_{12} } \right] , \quad \ell_{12} = {\rm lcm}(d_1, d_2)
\label{k36} \\
& = & \frac{1}{2} \left[ \frac{ X^2}{4 \ell_{12} } + X  \left( 1- 2   \left\{  \frac{ X}{ 2 \, \ell_{12} } \right\} \right)  
+ \ell_{12}  \left\{  \frac{ X}{ 2 \, \ell_{12} } \right\}  \left(     \left\{  \frac{ X}{ 2 \, \ell_{12} } \right\} - 1  \right)  \right]
\nonumber 
\earray
In this expression we keep the term $O(X^2)$ and drop the next ones that a numerical calculation show they  are
$O(X)$. We are led then to the approximation

\barray 
\sum_{m=1}^{X}  m \,  \alpha^2(m)  & \sim  &  \frac{\alpha^2}{4} X^2  \sum_{d_1, d_2 =1}^\infty  \frac{ \beta(d_1) \beta(d_2)}{ {\rm lcm} (d_1, d_2)}  + O(X)  
= \frac{\alpha^2}{2 \alpha_2} X^2 + O(X)
\label{k361}
\earray
where we used Eq. (\ref{k311}). See Fig. \ref{sumC22} (right) to verify numerically this equation. 
Finally, Eqs.  (\ref{k333}) and (\ref{k361}) implies  the asymptotic behavior of Eq. (\ref{k201})

\beq
\sum_{i, j=1}^{X/2} \alpha^2( 2 |i-j|)  \simeq  \frac{ \alpha^2}{2 \alpha_2} X^2-  \frac{X}{2}   \left( \log X \right)^2 + O(X) , \quad X \rightarrow \infty.
\label{k37}
\eeq
Fig. \ref{sumC22} (right) shows a numerical check of this relation, that exhibit  smaller fluctuations
as the previous approximations Eqs. (\ref{k333}) and (\ref{k361}). This is  probably due to the fact
that they contribute with opposite sign to Eq. (\ref{k201}). For larger values of $X$, not shown in 
Fig. \ref{sumC22} (right), the result becomes asymptotically exact.

\section*{Appendix A4: a toy  density matrix}

Let us suppose a density matrix of dimension $d$ of the form
\beq
\rho = \frac{1}{d} ( {\bf 1} + c  P), 
\label{bp1}
\eeq
where $P$ is the matrix 
\beq
P_{a,a'} = 
\left\{ \begin{array}{cc}
0,  &  {\rm if} \, a = a' \\
1,  & {\rm if} \, a \neq a' \\
\end{array}
\right. . 
\label{bp2}
\eeq
The eigenvalues of  $P$ are
\be
p_1 = d-1,  \qquad p_i = -1,  \;  i=2, \dots, d-1. 
\label{bp3}
\eeq
Hence the eigenvalues of $\rho$ are
\beq
\lambda_1 = \frac{ 1 + c(d-1)}{d}, \qquad   \lambda_i = \frac{1-c}{d}, \;  i=2, \dots, d-1. 
\label{bp4}
\eeq
The condition  $0 \leq \lambda_i \leq 1$ imposes that $0 \leq c \leq 1$. In the limit $d \gg 1$, the  purity
behaves as 
\beq
{\rm Tr} \rho^2 = \lambda_1^2 + (d-1) \lambda_2^2=   \left(  \frac{ 1 + c(d-1)}{d} \right)^2 +
(d-1) \left(  \frac{1-c}{d} \right)^2  \rightarrow  c^2 + \frac{ (1-c)^2}{d}  \simeq c^2, 
\label{bp5}
\eeq
where we used that $0 < c < 1$.  Hence the dominant behavior is controlled by the eigenvalue $\lambda_1 \simeq c$. 
We can also get this result from the square of $\rho$, 
\beq
\rho^2 = \frac{1}{d^2} ( {\bf 1} + 2 c  P + c^2  P^2) \Rightarrow
{\rm Tr} \rho^2 = \frac{ d + c^2  d(d-1)}{d^2} \rightarrow c^2, 
\label{bp6}
\eeq
which shows that the purity comes essentially from the square of the off-diagonal matrix $P$. 
In the same limit the von Neumann entropy behaves as 
\beq
S^{(1)} = - ( \lambda_1 \log \lambda_1 + (d-1) \lambda_2 \log \lambda_2 )  \rightarrow - c \log c + (1-c) \log \frac{d}{1-c}  \sim
(1-c) \log d. 
\label{bp7}
\eeq
Hence satisfies a volume law, with a coefficient that depends on the strength of the off diagonal term $P$.



\end{document}